\documentclass[aps,prd,twocolumn,superscriptaddress]{revtex4-1}

\usepackage{ulem}
\usepackage{amssymb}
\usepackage{amsmath}
\usepackage{dcolumn}
\usepackage{graphicx}
\usepackage{mathrsfs}
\usepackage{subfigure}
\usepackage{booktabs}
\usepackage{color}
\usepackage{docmute}
\usepackage{hyphenat}
\usepackage{CJKutf8}
\usepackage{multirow}
\usepackage{array}
\usepackage{threeparttable}
\usepackage[mathscr]{euscript}

\usepackage{graphicx}
\usepackage{subfigure}
\usepackage{tikz}
\usetikzlibrary{positioning}
\usepackage{xcolor}

\newcommand{\bea}{\begin{eqnarray}}
\newcommand{\eea}{\end{eqnarray}}
\newcommand{\beq}{\begin{equation}}
\newcommand{\eeq}{\end{equation}}
\newcommand{\nn}{\nonumber}
\def\/{\over}

\usepackage{url}
\usepackage[colorlinks]{hyperref}
\hypersetup{
	plainpages=true,
	breaklinks=true,
	hypertexnames=false,
	pageanchor=true,
	bookmarksnumbered=true,
	colorlinks=true,
	linkcolor={blue},
	citecolor={red},
	urlcolor={blue},
	anchorcolor={black}
}

\allowdisplaybreaks[4]

\begin{document}
\title{Spontaneous excitation of a centripetally accelerated atom coupled to electromagnetic vacuum fluctuations near a reflecting boundary}
\author{Yan Peng}
\affiliation{Department of Physics, Key Laboratory of Low Dimensional Quantum Structures and Quantum Control of Ministry of Education, and Hunan Research Center of the Basic Discipline for Quantum Effects and Quantum Technologies, Hunan Normal University, Changsha, Hunan 410081, China}
\author{Jiawei Hu}
\email[Corresponding author: ]{jwhu@hunnu.edu.cn}
\affiliation{Department of Physics, Key Laboratory of Low Dimensional Quantum Structures and Quantum Control of Ministry of Education, and Hunan Research Center of the Basic Discipline for Quantum Effects and Quantum Technologies, Hunan Normal University, Changsha, Hunan 410081, China}
\author{Hongwei Yu}
\email[Corresponding author: ]{hwyu@hunnu.edu.cn}
\affiliation{Department of Physics, Key Laboratory of Low Dimensional Quantum Structures and Quantum Control of Ministry of Education, and Hunan Research Center of the Basic Discipline for Quantum Effects and Quantum Technologies, Hunan Normal University, Changsha, Hunan 410081, China}

\begin{abstract}

We investigate the rate of change of the mean atomic energy for centripetally accelerated atoms interacting with electromagnetic vacuum fluctuations near a reflecting boundary, using the Dalibard–Dupont-Roc–Cohen-Tannoudji formalism. The distinct contributions from vacuum fluctuations and radiation reaction are analyzed separately. Our results reveal that, when the centripetal acceleration significantly exceeds the characteristic acceleration set by the atomic transition frequency, vacuum fluctuations dominates over radiation reaction, irrespective of the atom-boundary distance and the atomic polarization. 
In the near-zone regime, where the atom-boundary distance is much smaller than  both the characteristic length  associated with the acceleration and the transition wavelength of the atom, the boundary introduces substantial corrections to the rate of change of the mean atomic energy.
These corrections are comparable in magnitude to those in free space and exhibit strong dependence on the atomic polarization.
Remarkably, in the intermediate and far regions, contributions stemming from the combined effects of the boundary and acceleration can become the leading and subleading terms, respectively. An acceleration-independent term also arises from their interplay.  
These findings highlight the significant  interplay between acceleration and the presence of a boundary  in shaping atomic radiative properties and may have potential implications for experimentally probing the circular Unruh effect.

\end{abstract}

\maketitle

\section{Introduction}

Spontaneous emission, a fundamental radiative property of atoms, has long captivated researchers. Its underlying physical mechanisms have been attributed to vacuum fluctuations~\cite{Welton48,Compagno83}, radiation reaction~\cite{Ackerhalt73}, or a combination of both~\cite{Senitzky73,Milonni75}. The ambiguity arises fundamentally from differing choices in the ordering of the atomic and field operators, leading to uncertainty in the contributions of vacuum fluctuations and radiation reaction. 
The issue was resolved  by Dalibard, Dupont-Roc, and Cohen-Tannoudji (DDC) \cite{JJCJ82,JJCJ84}, who proposed a symmetric ordering of the atomic and field operators. Their approach ensures that the rates of change of atomic observables due to vacuum fluctuations and radiation reaction are separately Hermitian, thus granting each independent physical significance.

The DDC formalism has been widely employed to study the radiative properties of atoms moving along various trajectories and  interacting with different kinds of quantum fields~\cite{JR94,JR95,JRM95,Passante98,Zhu06,Rizzuto07,Rizzuto09,Zhu09,Zhu10,Zhou12,Li14,Yu07,Yu071,Zhou10,Yu19}. 
According to this formalism, the stability of inertial atoms in the ground state is due to the exact cancellation of the contribution of vacuum fluctuations and radiation reaction to the rate of change of the mean atomic energy. However, for uniformly accelerated atoms in the ground state, this delicate  balance is disrupted, enabling spontaneous excitation \cite{JR94}. This phenomenon can be viewed as a manifestation of the Unruh effect~\cite{Fulling73,Davies75,Unruh76}, which attests that a uniformly accelerated observer perceives the vacuum seen by an inertial observer as a thermal bath at a temperature proportional to the observer's proper acceleration. The Unruh effect holds profound theoretical significance, both as a stand-alone prediction and for its deep connection to the Hawking effect in black holes, as suggested by the equivalence principle. However, detecting the Unruh effect would require extraordinarily high accelerations due to the fact that the effect is extremely small at commonly achievable accelerations, presenting a significant challenge with current technological capabilities.

Apart from uniform linear acceleration, another common type of acceleration is uniform centripetal acceleration. 
Compared to uniform linear acceleration, greater acceleration can be achieved more easily and atoms under centripetal acceleration are confined to a localized region of space. Consequently, there has been significant interest in studying field quantization in rotating frames and understanding the effects of centripetal acceleration on the radiative properties of atoms~\cite{Letaw80,Letaw81,Takagi84,Hacyan86,Bell87,Kim87,Bell83,Rogers88,Davies1996,Unruh98,Lorenci2000,Rosu2005,Biermann2020,Jin14}.
It has been found that, unlike in the case of uniform linear acceleration, the radiation perceived by rotating observers is inherently  nonthermal. Using the DDC formalism, it has been shown that, similar to atoms under uniform linear acceleration, the balance between vacuum fluctuations and radiation reaction is also disrupted for centripetally accelerated atoms, making spontaneous excitation possible~\cite{JRM95,Jin14}. 
Furthermore, in contrast to the linear acceleration case,   the contribution of vacuum fluctuations for centripetally accelerated atoms is not proportional to a Planckian factor. Instead, it is proportional to a non-Planckian exponential term, indicating that that the radiation perceived by centripetally orbiting observers deviates  from a thermal spectrum.

The examples above show that  vacuum fluctuations perceived by noninertial observers, such as those under linear or centripetal acceleration, are significantly different from those observed by inertial observers. Furthermore, vacuum fluctuations can also be significantly modified by the presence of boundaries, 
as exemplified  by the well-known Casimir force between neutral conducting plates. This force arises from boundary-induced modifications to the vacuum~\cite{Casimir48} and underscores how such changes can affect the properties of vacuum quantum fields. 
 Boundary-induced modifications have a substantial impact on the radiative properties of atoms, making them markedly different from those in free space.
For instance,  in a high-$Q$ cavity,  the spontaneous emission rate of an atom can be substantially amplified, an effect commonly known as the Purcell effect~\cite{Purcell46,Purcell461}.  This naturally leads to the question: how do the radiative properties of noninertial atoms behave near a boundary?

When modeling the atom-field interaction as a monopole interaction with vacuum scalar field fluctuations, it is revealed that the rate of change of the mean atomic energy for uniformly accelerated atoms near a reflecting boundary oscillates as a function of the atom-boundary distance  \cite{Yu061}. This rate decreases as the atom approaches the boundary   and eventually vanishes on the  boundary.  
 For the more realistic case of the dipole interaction with fluctuating electromagnetic fields, the rate of change of the mean atomic energy in free space includes a nonthermal correction~\cite{Yu062}. The presence of a reflecting boundary can either amplify or suppress this correction, depending on the atom's distance from the boundary~\cite{Yu06}. Furthermore, with a reflecting boundary, the rate of energy change is found to depend on the atomic polarization~\cite{Yu06}. 
Since the vacuum perceived by a centripetally accelerated atom is qualitatively distinct from that perceived by a uniformly accelerated one, it is especially intriguing to explore how the radiative properties of centripetally accelerated atoms are modified by a reflecting boundary compared to those in free space~\cite{Jin14}.

For centripetally accelerated atoms, previous studies have investigated radiative properties in more complex boundary geometries \cite{Lochan20,Arya23,guo24,Levin93}, revealing a variety of interesting phenomena. For instance, it has been shown that, in a cylindrical cavity, a critical centripetal acceleration is required for spontaneous excitation to occur \cite{Levin93,guo24}. In contrast, for cavities characterized by a Lorentzian spectral density, no  such threshold exists~\cite{Lochan20}. However, both the transition rate~\cite{Lochan20} and the Lamb shift~\cite{Arya23} can be significantly enhanced compared to those in free space.  Despite these findings, the complexity of the boundary geometries leaves unresolved whether the cavity and centripetal acceleration influence transition rates independently or interact in nontrivial, possibly synergistic, ways.

In this paper, we investigate the rate of change of the mean atomic energy for centripetally accelerated atoms coupled to fluctuating electromagnetic fields in a vacuum near a reflecting boundary using the DDC approach. 
The simplicity of the boundary geometry  enables us to disentangle the individual roles of the  boundary and centripetal acceleration in modifying atomic transition rates. More importantly, the DDC approach allows for a clear separation of the effects due to vacuum fluctuations and radiation reaction to the rate of change of atomic energy. As a result, this work provides a  more transparent understanding of how the interplay between centripetal acceleration and the presence of a boundary influences the radiative properties of atoms through both vacuum fluctuations and radiation reaction channels.   
The paper is structured as follows. In Sec.~\ref{sec2}, we present the general formalism for calculating the rate of change of the mean atomic energy of a multilevel atom interacting with electromagnetic vacuum fluctuations in the presence of a reflecting boundary using the DDC approach. 
In Sec.~\ref{sec3}, we provide a detailed analysis of the contributions of vacuum fluctuations and radiation reaction to the rate of change of the mean atomic energy for a centripetally accelerated atom.  
Finally, we summarize in Sec.~\ref{sec5}. This paper is formulated in natural units $\hbar=c=k_{B}=\epsilon_0=1$, with $\hbar$ representing the reduced Planck constant, $c$ the speed of light, $k_{B}$ the Boltzmann constant, and $\epsilon_0$ the vacuum permittivity.

\section{The basic formalism} \label{sec2}
In this section, we briefly review the DDC method established in Refs.~\cite{JJCJ82,JJCJ84}. We consider a multilevel atom coupled to electromagnetic vacuum fluctuations near a reflecting boundary. In the Heisenberg picture, the Hamiltonian governing the time evolution of the atom with respect to the proper time $\tau$ is given by
\begin{eqnarray}\label{HA}
H_A(\tau)=\sum_n\omega_n\sigma_{nn}(\tau)\;,
\end{eqnarray}
where $\omega_n$ is the eigenenergy corresponding to the energy eigenstate $|n\rangle$, and $\sigma_{nn}(\tau)=|n \rangle \langle n|$. The Hamiltonian of the electromagnetic field is
\begin{eqnarray}\label{HF}
H_F(\tau)=\sum_k \omega_{\vec{k}} a_{\vec{k}}^\dag a_{\vec{k
}}{dt\/d \tau}\;.
\end{eqnarray}
Here $\vec{k}$ represents the wave vector and polarization of the field modes, $\omega_{\vec{k}}$ is the frequency of field mode $\vec{k}$, and $a_{\vec{k}}^\dag$, $ a_{\vec{k}}$ are the creation and annihilation operators, respectively. 
The interaction Hamiltonian for the atom-field coupling is expressed as
\begin{align}\label{HI}
H_I(\tau)&=-e\textbf{ r}(\tau) \cdot
\textbf{E}(x(\tau))\nonumber\\
&=-e\sum_{mn}\textbf{r}_{mn}\cdot
\textbf{E}(x(\tau))\;\sigma_{mn}(\tau)\;,
\end{align}
where $e\textbf{r}$ is the atomic electric dipole moment, with $e$ being the electron charge and $x(\tau)$ denoting the space-time coordinates of the atom.

The Heisenberg equations of motion for the dynamical variables of the atom and the electromagnetic field can be derived from the total Hamiltonian $H=H_A+H_F+H_I$. 
The solutions to these equations can be divided into two parts: the free part, which exists even when the atom and field are decoupled, and the source part, which arises from the interaction between the atom and the field. 
In order to separate the contributions of vacuum fluctuations and radiation reaction to the rate of change of a physical variable, we choose a symmetric ordering of the atomic and field variables, which is a key step in the DDC method. 
We assume that the initial states of the electromagnetic field and the atom are the vacuum state $|0\rangle$ and the atomic state $|b\rangle$, respectively. 
Since we are interested in the spontaneous emission and excitation of the atom, we focus on the mean atomic energy $\langle H_A\rangle$. The contributions of vacuum fluctuations $(\rm VF)$ and radiation reaction $(\rm RR)$ to the rate of change of the mean atomic energy can be expressed as
\begin{align}
   & {\biggl\langle
{dH_A(\tau)\/d\tau}\biggr\rangle_{\rm VF}}=\nonumber\\
&~~~~~~~~~~2ie^2\int_{\tau_0}^\tau
d\tau' C_{ij}^F(x(\tau),x(\tau')){d\/d\tau}(\chi_{ij}^A)_b
(\tau,\tau')\label{hvf0}\;
\end{align}
and 
\begin{align}
&{\biggl\langle
{dH_A(\tau)\/d\tau}\biggr\rangle_{\rm RR}}=\nonumber\\
&~~~~~~~~~~2ie^2\int_{\tau_0}^\tau
d\tau' \chi_{ij}^F(x(\tau),x(\tau')){d\/d\tau}(C_{ij}^A)_b
(\tau,\tau')\label{hrr0}\;,
\end{align}
respectively.  
Here, $\langle {\cal O} \rangle$ denotes the expectation value of the operator ${\cal O}$ with respect to the state $|b,0\rangle$. 
\begin{align}
C_{ij}^F&={1\/2}\langle0|\{E_i^f(x(\tau)),E_j^f(x(\tau'))\}|0\rangle\label{cf}\;,\\
\chi_{ij}^F&={1\/2}\langle0|[E_i^f(x(\tau)),E_j^f(x(\tau'))]|0\rangle\;\label{xf}
\end{align}
are the statistical functions of the field, and 
\begin{align}
(C_{ij}^A)_b(\tau,\tau')&={1\/2}\langle
b|\{r_i^f(\tau),r_j^f(\tau')\}|b\rangle\;,\label{CA}\\
(\chi_{ij}^A)_b(\tau,\tau')&={1\/2}\langle
b|[r_i^f(\tau),r_j^f(\tau')]|b\rangle\;\label{XA}
\end{align}
are the statistical functions of the atom. 
Taking the solution of the atomic operator $r_i^f(\tau)$ into Eqs.~\eqref{CA} and \eqref{XA}, the explicit forms of the statistical functions of the atom are found to be
\begin{align}
(C_{ij}^A)_b(\tau,\tau')=&{1\/2}\sum_d[\langle
b|r_i(0)|d\rangle\langle d|r_j(0)|b\rangle
e^{i\omega_{bd}(\tau-\tau')}\nonumber\\
&+\langle b|r_j(0)|d\rangle\langle
d|r_i(0)|b\rangle e^{-i\omega_{bd}(\tau-\tau')}]\label{ca}\;
\end{align}
and
\begin{align}
(\chi_{ij}^A)_b(\tau,\tau')=&{1\/2}\sum_d[\langle
b|r_i(0)|d\rangle\langle d|r_j(0)|b\rangle
e^{i\omega_{bd}(\tau-\tau')}\nonumber\\
&-\langle b|r_j(0)|d\rangle\langle
d|r_i(0)|b\rangle e^{-i\omega_{bd}(\tau-\tau')}]\label{xa}\;.
\end{align}
Here, $\omega_{bd}=\omega_b-\omega_d$, and the summation runs over a complete set of atomic states.

In order to get the statistical functions of the electromagnetic field, we begin with the two-point function of the vector potential $\langle0| A_\mu(x(\tau)) A_{\nu}(x(\tau'))|0\rangle=D_{\mu\nu}^{\rm free}(x,x')+D_{\mu\nu}^{\rm bnd}(x,x')$, where ``free'' denotes the contribution in free space and ``bnd'' represents the correction term introduced by the boundary. In the laboratory frame, the two-point functions of the vector potential in the Feynman gauge 
$D_{\mu\nu}^{\rm free}(x,x')$ and $D_{\mu\nu}^{\rm bnd}(x,x')$ 
can be obtained using the method of images as
\begin{align}
&D_{\mu\nu}^{\rm free}(x,x')=\nonumber\\&\frac{\eta_{\mu\nu}}{{4\pi^2
[(x-x^\prime)^2+(y-y^\prime)^2+(z-z^\prime)^2-(t-t^\prime-i\varepsilon)^2]}}\;
\end{align}
and
\begin{align}
&D^{\rm bnd}_{\mu\nu}(x,x')=\nonumber\\&-{\frac{\eta_{\mu\nu}-2n_\mu n_\nu}{4\pi^2
[(x-x^\prime)^2+(y-y^\prime)^2+(z+z^\prime)^2-(t-t^\prime-i\varepsilon)^2]}}\;,
\end{align}
respectively, 
where $\eta_{\mu \nu}=\rm {diag}(-1,1,1,1)$ and $n_{\mu}=(0,0,0,1)$. 
Recall that $E_i=A_{0,i}-A_{i,0}$, so the total two-point function of the electric field can also be expressed as the sum of the free-space part and a correction term due to the presence of the boundary
\begin{align}
&\langle0| {E}(x){E}(x')|0\rangle=\nonumber\\&\langle0|
{E}(x){E}(x')|0\rangle_{\rm free}+\langle0|
{E}(x){E}(x')|0\rangle_{\rm bnd}\;,
\end{align}
where 
\begin{align}
&\langle0|E_m(x(\tau))E_n(x(\tau'))|0\rangle_{\rm free}
=\nonumber \\
&\frac{-\partial_0\partial'_0\delta_{mn}+\partial_m\partial'_n}{4\pi^2[(x-x')^2+(y-y')^2+(z-z')^2-(t-t'-i\varepsilon)^2]}\;
\end{align}
and
\begin{align}
&\langle0|
E_m(x(\tau))E_n(x(\tau'))|0\rangle_{\rm bnd}=\nonumber\\
&{\frac{\,(\delta_{mn}-2n_m n_n)\,\partial
_0\partial_0^\prime-\partial_m\partial_n^\prime\,}{4\pi^2[(x-x^\prime)^2+(y-y^\prime)^
2+(z+z^\prime)^2-(t-t^\prime-i\varepsilon)^2]}}\;.
\end{align}
Here, $\varepsilon\rightarrow+0$, and $\partial^\prime$ denotes the derivative with respect to $x^\prime$. The statistical functions of the electromagnetic field Eqs.~\eqref{cf} and \eqref{xf} can then be calculated using the two-point functions above.

\section{Spontaneous excitation of a centripetally accelerated atom near a reflecting boundary}\label{sec3}
\begin{figure}[ht]
\includegraphics[scale=0.2]{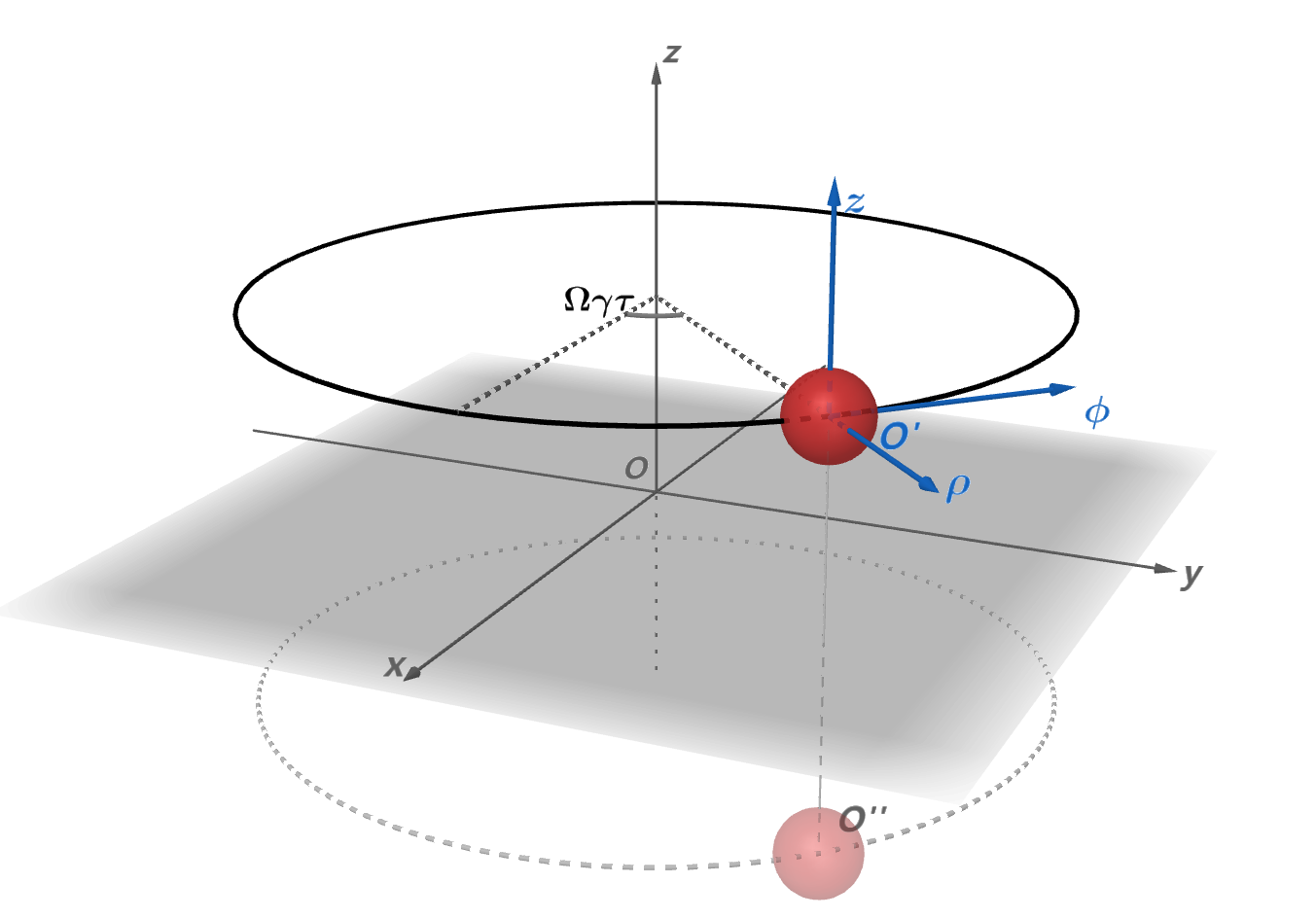}\vspace{0.0cm}
\caption{\label{Coordinate} 
Centripetally accelerated atom located near a reflecting boundary, with its orbit aligned parallel to the boundary. 
The laboratory frame is represented in Cartesian coordinates $(t,x,y,z)$, while the atom's proper frame is described in cylindrical coordinates $(t,\rho,\phi,z)$.}
\end{figure}

Now, we consider a centripetally accelerated atom coupled to fluctuating electromagnetic fields near a reflecting boundary, as shown in Fig.~{\ref{Coordinate}}. 
The trajectory of the atom can be described as
\begin{align}
t(\tau)=\gamma \tau\;,~~x(\tau)=R\cos\Omega\gamma\tau\;,\nonumber\\
y(\tau)=R\sin\Omega\gamma \tau\;,~~z(\tau)=L\;.
\end{align}
Here, $R$ is the orbital radius, $\Omega$ is the rotational angular velocity, $L$ is the distance between the atom and the boundary, and $\gamma=1/\sqrt{1-\Omega^2 R^2}$ is the Lorentz factor. 

As we calculate the rate of change of the mean atomic energy in the proper frame of the atom, the electromagnetic field correlation functions in this frame are required. These functions can be obtained by a Lorentz transformation together with a rotational transformation. The detailed derivations and results are shown in Appendix \ref{correlation-function}. Here, we consider the ultrarelativistic limit, i.e., the limit when the linear velocity $v=R\Omega\to 1$. The explicit expressions for the field correlation functions $G_{ij}(u)$ in this limit, which are functions of $u=\tau-\tau^\prime$ as they are invariant under temporal translations, are shown in Appendix~\ref{correlation-function}. See Eqs.~\eqref{Gzz1}$\sim$\eqref{zrho1} for details. 
Then, the contributions of vacuum fluctuations and radiation reaction to the rate of change of the mean atomic energy can be derived with the help of Eqs. (\ref{hvf0}) and (\ref{hrr0}) as 
\begin{align}
&{\biggl\langle
{\frac{dH_A(\tau)}{d\tau}}\biggr\rangle_{\rm VF}}=-\frac{e^2}{2} \sum_d \langle b|r_i(0)|d\rangle\langle d|r_j(0)|b\rangle\;\omega_{bd}  \nonumber\\
&\int_{-\infty}^\infty du  \left[G_{ij}(u-i \epsilon )+G_{ij}(u+i \epsilon )\right] 
e^{i \omega _{bd} u}\;\label{ijVF}
 \end{align}
 and
 \begin{align}
&{\biggl\langle
{\frac{dH_A(\tau)}{d\tau}}\biggr\rangle_{\rm RR}}=-\frac{e^2}{2} \sum_d \langle b|r_i(0)|d\rangle\langle d|r_j(0)|b\rangle \;\omega_{bd} \nonumber\\
& \int_{-\infty}^\infty du  \left[G_{ij}(u-i \epsilon )-G_{ij}(u+i \epsilon )\right]
e^{i \omega _{bd} u}\;,\label{ijRR}
 \end{align}
respectively.  
The indices $i,j={\rho,\phi,z}$, and repeated indices of $i,j$ means summation. For simplicity, 
we define $\bar{\mathcal{R}}_{ij}\equiv \langle b|r_i(0)|d\rangle \langle d|r_j(0)|b\rangle$ and assume it to be real, so $\bar{\mathcal{R}}_{ij} = \bar{\mathcal{R}}_{ji}$. 
Since the general results are too intricate and not particularly enlightening, we opt not to explicitly display them. Instead, we focus on deriving meaningful insights from the complex general expressions by examining specific asymptotic regions. This approach enables us to emphasize the key physical effects stemming from centripetal acceleration, boundary modifications, and atomic polarization.
By simplifying the analysis to these targeted regions, we can effectively showcase the interplay of these factors and provide a clearer understanding of their individual and combined influences on the rate of change of the mean atomic energy.

\subsection{Far region with $L \gg 1/a$ and $1/|\omega_{bd}|$}

As the atom-boundary distance approaches infinity ($L \rightarrow \infty$), the contributions of vacuum fluctuations and radiation reaction to the rate of change of the mean atomic energy reduce to their corresponding free-space forms, as expected. The explicit expressions are given in Eqs.~\eqref{free-space-vf} and \eqref{free-space-rr} in Appendix~\ref{free-average-rate}. Notably, the contributions differ depending on the atomic polarization, exhibiting anisotropic behavior. This anisotropy stands in contrast to the earlier results for centripetally accelerated atoms in free space, reported in Refs.~\cite{Jin14} and \cite{Jin141}, where the contributions from different atomic polarizations were found to be identical. The discrepancy in Refs.~\cite{Jin14} and \cite{Jin141} arose from errors in the Lorentz transformation of the electromagnetic field.

To further examine the interplay between centripetal acceleration and boundary effects on the rate of change of the mean atomic energy, we divide the far region, where the atom-boundary distance $L$ is much larger than both the characteristic length determined by the acceleration $1/a$ and the transition wavelength of the atom $1/|\omega_{bd}|$, into two subregions: $1/a \ll 1/|\omega_{bd}| \ll L$ and $ 1/|\omega_{bd}| \ll1/a \ll L$.

\subsubsection{Far region with $1/a \ll 1/|\omega_{bd}| \ll L$}
When the atomic transition wavelength $1/|\omega_{bd}|$ is much smaller than the atom-boundary distance $L$, but much larger than the characteristic length determined by the centripetal acceleration $1/a$, the contribution of vacuum fluctuations to the rate of change of the mean atomic energy for centripetally accelerated atoms near the boundary is given by
\begin{widetext}
    \begin{align}
   {\biggl\langle {\frac{dH_A(\tau)}{d\tau}}\biggr\rangle}_{\rm VF} \approx
&{-\frac{ e^2}{3\pi}
} \left(\sum_{\omega_b>\omega_d}-\sum_{\omega_b<\omega_d}\right)\omega_{bd}^4
\,\bigg \{  \frac{a^3 \left(58 \bar{\mathcal{R}}_{\rho \rho}+55 \bar{\mathcal{R}}_{\phi\phi}+13 \bar{\mathcal{R}}_{zz}\right)}{32 \sqrt{3} |\omega_{bd}|^3}+\frac{\sqrt{3} a \left(15 \bar{\mathcal{R}}_{\phi\phi}+14 \bar{\mathcal{R}}_{\rho \rho}+5 \bar{\mathcal{R}}_{zz}\right)}{16 |\omega_{bd}|}\nn\\
&-\frac{9 \sqrt[4]{3} \sqrt{a L} \left(2 \bar{\mathcal{R}}_{z \rho}-\sqrt{3}\bar{\mathcal{R}}_{\rho \rho}+\sqrt{3} \bar{\mathcal{R}}_{zz}\right)}{128 L^3 |\omega_{bd}|^3}\;\bigg \}\;,\label{Lfarvf2}
\end{align}
and that of radiation reaction is
\begin{align}
   {\biggl\langle {\frac{dH_A(\tau)}{d\tau}}\biggr\rangle}_{\rm RR} \approx
&{-\frac{ e^2}{3\pi}
} \left(\sum_{\omega_b>\omega_d}+\sum_{\omega_b<\omega_d}\right)\omega_{bd}^4
\,\bigg \{ \frac{a^2 \left(4 \bar{\mathcal{R}}_{\rho \rho}+4 \bar{\mathcal{R}}_{\phi\phi}+\bar{\mathcal{R}}_{zz}\right)}{2 |\omega_{bd}|^2}+\frac{1}{2} \left(\bar{\mathcal{R}}_{\rho \rho}+\bar{\mathcal{R}}_{\phi\phi}+\bar{\mathcal{R}}_{zz}\right)\;\bigg \}\;.\label{Lfarrr2}
\end{align}
Then, the total rate of change of the atomic energy is
\begin{align}
{\biggl\langle {\frac{dH_A(\tau)}{d\tau}}\biggr\rangle}_{\rm tot} \approx
&{-\frac{ e^2}{3\pi}
} \sum_{\omega_b>\omega_d}\omega_{bd}^4
\,\bigg \{ \frac{a^3 \left(58 \bar{\mathcal{R}}_{\rho \rho}+55 \bar{\mathcal{R}}_{\phi\phi}+13 \bar{\mathcal{R}}_{zz}\right)}{32 \sqrt{3} |\omega_{bd}|^3}+\frac{a^2 \left(4 \bar{\mathcal{R}}_{\rho \rho}+4 \bar{\mathcal{R}}_{\phi\phi}+\bar{\mathcal{R}}_{zz}\right)}{2 |\omega_{bd}|^2}\nn\\
&+\frac{\sqrt{3} a \left(15 \bar{\mathcal{R}}_{\phi\phi}+14 \bar{\mathcal{R}}_{\rho \rho}+5 \bar{\mathcal{R}}_{zz}\right)}{16 |\omega_{bd}|}+\frac{1}{2} \left(\bar{\mathcal{R}}_{\rho \rho}+\bar{\mathcal{R}}_{\phi\phi}+\bar{\mathcal{R}}_{zz}\right)\nn\\
&-\frac{9 \sqrt[4]{3} \sqrt{a L} \left(2 \bar{\mathcal{R}}_{z \rho}-\sqrt{3}\bar{\mathcal{R}}_{\rho \rho}+\sqrt{3} \bar{\mathcal{R}}_{zz}\right)}{128 L^3 |\omega_{bd}|^3}\;\bigg \}\nonumber\\
&+\frac{ e^2}{3\pi}\sum_{\omega_b<\omega_d} \omega_{bd}^4
\, \bigg \{\frac{a^3 \left(58 \bar{\mathcal{R}}_{\rho \rho}+55 \bar{\mathcal{R}}_{\phi\phi}+13 \bar{\mathcal{R}}_{zz}\right)}{32 \sqrt{3} |\omega_{bd}|^3}-\frac{a^2 \left(4 \bar{\mathcal{R}}_{\rho \rho}+4 \bar{\mathcal{R}}_{\phi\phi}+\bar{\mathcal{R}}_{zz}\right)}{2 |\omega_{bd}|^2}\nn\\
&+\frac{\sqrt{3} a \left(15 \bar{\mathcal{R}}_{\phi\phi}+14 \bar{\mathcal{R}}_{\rho \rho}+5 \bar{\mathcal{R}}_{zz}\right)}{16 |\omega_{bd}|}-\frac{1}{2} \left(\bar{\mathcal{R}}_{\rho \rho}+\bar{\mathcal{R}}_{\phi\phi}+\bar{\mathcal{R}}_{zz}\right)\nn\\
&-\frac{9 \sqrt[4]{3} \sqrt{a L} \left(2 \bar{\mathcal{R}}_{z \rho}-\sqrt{3}\bar{\mathcal{R}}_{\rho \rho}+\sqrt{3} \bar{\mathcal{R}}_{zz}\right)}{128 L^3 |\omega_{bd}|^3}
\;\bigg \}\;.\label{Lfartot2}
\end{align}
\end{widetext}

\emph{1.}  According to Eq.~(\ref{Lfartot2}), for a centripetally accelerated atom in the ground state ($\omega_b<\omega_d$), the contributions of vacuum fluctuations and radiation reaction to the rate of change of the mean atomic energy do not cancel as they do in the inertial case. As a result, a centripetally accelerated ground-state atom can spontaneously become excited, which is a manifestation of the circular Unruh effect. This demonstrates how noninertial motion breaks the balance of contributions that characterize the inertial case.  
In this regime, the leading contribution of vacuum fluctuations to the rate of change of the mean atomic energy scales as $a^3/|\omega_{bd}|^3$, which significantly exceeds the radiation reaction contribution scaling as $a^2/|\omega_{bd}|^2$, given that  $a/|\omega_{bd}| \gg 1$. As a result, vacuum fluctuations dominate the rate of change of the mean atomic energy, irrespective of whether the atom is in the ground or excited state.  

\emph{2.}  For comparison, we present the contributions of vacuum fluctuations and radiation reaction to the rate of change of the mean atomic energy for inertial atoms in the regime $1/|\omega_{bd}| \ll L$ as follows: 
\begin{widetext}
\begin{align}
{\biggl\langle {\frac{dH_A(\tau)}{d\tau}}\biggr\rangle}_{\rm VF}^{\rm in} \approx
&{-\frac{ e^2}{3\pi}
} \left(\sum_{\omega_b>\omega_d}-\sum_{\omega_b<\omega_d}\right)\omega_{bd}^4
\,\bigg [\frac{\bar{\mathcal{R}}_{\rho \rho}+\bar{\mathcal{R}}_{\phi\phi }+\bar{\mathcal{R}}_{zz}}{2} -\frac{3 \sin (2 L |\omega_{bd}|) \left(\bar{\mathcal{R}}_{\rho \rho}+\bar{\mathcal{R}}_{\phi\phi }\right)}{8 L |\omega_{bd}|}\nn\\
&-\frac{3 [2 L |\omega_{bd}| \cos (2 L |\omega_{bd}|)-\sin (2 L |\omega_{bd}|)]  \left(2 \bar{\mathcal{R}}_{zz}+\bar{\mathcal{R}}_{\rho \rho}+\bar{\mathcal{R}}_{\phi\phi }\right)}{32 L^3 |\omega_{bd}|^3}\;\bigg ]\;\label{in-vf}
\end{align}
and
\begin{align}
{\biggl\langle {\frac{dH_A(\tau)}{d\tau}}\biggr\rangle}_{\rm RR}^{\rm in} \approx
&{-\frac{ e^2}{3\pi}
} \left(\sum_{\omega_b>\omega_d}+\sum_{\omega_b<\omega_d}\right)\omega_{bd}^4
\,\bigg [\frac{\bar{\mathcal{R}}_{\rho \rho}+\bar{\mathcal{R}}_{\phi\phi }+\bar{\mathcal{R}}_{zz}}{2} -\frac{3 \sin (2 L |\omega_{bd}|) \left(\bar{\mathcal{R}}_{\rho \rho}+\bar{\mathcal{R}}_{\phi\phi }\right)}{8 L |\omega_{bd}|}\nn\\
&-\frac{3 [2 L |\omega_{bd}| \cos (2 L |\omega_{bd}|)-\sin (2 L |\omega_{bd}|)]  \left(2 \bar{\mathcal{R}}_{zz}+\bar{\mathcal{R}}_{\rho \rho}+\bar{\mathcal{R}}_{\phi\phi }\right)}{32 L^3 |\omega_{bd}|^3}\;\bigg ]\;.\label{in-rr}
\end{align}
\end{widetext}
By comparing Eqs.~\eqref{Lfarvf2} and \eqref{Lfarrr2} with Eqs.~\eqref{in-vf} and\eqref{in-rr}, it is evident that acceleration has a substantial impact on the rate of change of atomic energy, with acceleration-related terms dominating in this region.

\emph{3.} Comparing the results in this region Eqs.~(\ref{Lfarvf2}) and (\ref{Lfarrr2}) with those in free space in the large acceleration limit $a/ |\omega_{bd}|\gg 1$ [Eqs.~(\ref{freevf}) and (\ref{freerr})], we find that the leading contributions from radiation reaction in the two cases are identical. However, an additional correction term proportional to $\frac{\sqrt{a L}}{L^3|\omega _{bd}|^3}$ arises in the vacuum fluctuation contribution due to the combined effects of acceleration and the boundary. 
This term constitutes a correction to the free-space result and decreases monotonically as the atom-boundary distance increases.

\emph{4.} For both ground-state and excited-state atoms, Eq.~(\ref{Lfartot2}) shows that transverse and axial polarizations, corresponding to directions perpendicular and parallel to the axis of rotation, respectively, contribute comparably to the rate of change of the mean atomic energy in this regime. However, in the boundary-induced correction, polarization along the $\phi$-direction does not contribute to the leading term, and its overall contribution is significantly smaller than that of polarizations in the other directions.

\subsubsection{Far region with $ 1/|\omega_{bd}| \ll1/a \ll L$}
When the characteristic length determined by the acceleration $1/a$ is much smaller than the distance between the atom and the boundary $L$ and much larger than the atomic transition wavelength $ 1/|\omega_{bd}|$, the contributions of vacuum fluctuations and radiation reaction to the rate of change of the mean atomic energy can be simplified and derived using Eqs.~(\ref{ijVF}) and (\ref{ijRR}) as 
\begin{widetext}
\begin{align}
{\biggl\langle {\frac{dH_A(\tau)}{d\tau}}\biggr\rangle}_{\rm VF} \approx
&{-\frac{ e^2}{3\pi}
} \left(\sum_{\omega_b>\omega_d}-\sum_{\omega_b<\omega_d}\right)\omega_{bd}^4
\,\bigg \{\frac{\bar{\mathcal{R}}_{\rho \rho}+\bar{\mathcal{R}}_{\phi\phi }+\bar{\mathcal{R}}_{zz}}{2} + \frac{a^2 \left(4 \bar{\mathcal{R}}_{\rho \rho}+4\bar{\mathcal{R}}_{\phi\phi }+\bar{\mathcal{R}}_{zz}\right)}{2 |\omega_{bd}|^2} +e^{-\frac{2 \sqrt{3} |\omega_{bd}|}{a}}\times\nonumber\\
&\left[\frac{\sqrt{3} a \left(4 \bar{\mathcal{R}}_{\rho \rho}+3 \bar{\mathcal{R}}_{\phi\phi }+\bar{\mathcal{R}}_{zz}\right)}{8 |\omega_{bd}|}+\frac{a^2 \left(26 \bar{\mathcal{R}}_{\rho \rho}+23 \bar{\mathcal{R}}_{\phi\phi }+5 \bar{\mathcal{R}}_{zz}\right)}{16 |\omega_{bd}|^2}+\frac{a^3 \left(58 \bar{\mathcal{R}}_{\rho \rho}+55 \bar{\mathcal{R}}_{\phi\phi }+13 \bar{\mathcal{R}}_{zz}\right)}{32 \sqrt{3} |\omega_{bd}|^3}\right]\nonumber\\
&+\frac{3 a L \left[\bar{\mathcal{R}}_{\rho \rho} \left(p_{\rho }+q_{\rho }\right)-\bar{\mathcal{R}}_{zz} \left(p_z+q_z\right)+2\sqrt{3} \bar{\mathcal{R}}_{z \rho } \left(q_{z\rho }-p_{z\rho }\right)\right]+3\sqrt{3} \bar{\mathcal{R}}_{\phi\phi } \left(p_{\phi }-q_{\phi }\right)}{64 a L^3 |\omega _{bd}|^2}\;\bigg \}\;\label{Lfarvf}
\end{align}
and
\begin{align}
{\biggl\langle {\frac{dH_A(\tau)}{d\tau}}\biggr\rangle}_{\rm RR} \approx
&{-\frac{ e^2}{3\pi}
}\left(\sum_{\omega_b>\omega_d}+\sum_{\omega_b<\omega_d}\right)\omega_{bd}^4
\bigg [\frac{\bar{\mathcal{R}}_{\rho \rho}+\bar{\mathcal{R}}_{\phi\phi }+\bar{\mathcal{R}}_{zz}}{2}+\frac{a^2 \left(4 \bar{\mathcal{R}}_{\rho \rho}+4\bar{\mathcal{R}}_{\phi\phi }+\bar{\mathcal{R}}_{zz}\right)}{2 |\omega _{bd}|^2} \nonumber\\
&+\frac{3 a L \left( \bar{\mathcal{R}}_{\rho \rho}p_{\rho }-\bar{\mathcal{R}}_{zz} p_z -2\sqrt{3} \bar{\mathcal{R}}_{z \rho } p_{z\rho }\right)+3 \sqrt{3} \bar{\mathcal{R}}_{\phi\phi } p_{\phi }}{64 a L^3 |\omega _{bd}|^2}\bigg]\;,\label{Lfarrr}
\end{align}
respectively. 
By adding Eqs. \eqref{Lfarvf} and \eqref{Lfarrr} together, one obtains the total rate of change of the mean atomic energy as 
\begin{align}
{\biggl\langle {\frac{dH_A(\tau)}{d\tau}}\biggr\rangle}_{\rm tot} \approx
&{-\frac{ e^2}{3\pi}
} \sum_{\omega_b>\omega_d}\omega_{bd}^4
\,\bigg \{\bar{\mathcal{R}}_{\rho \rho}+\bar{\mathcal{R}}_{\phi\phi }+\bar{\mathcal{R}}_{zz}+ \frac{a^2 \left(4 \bar{\mathcal{R}}_{\rho \rho}+4\bar{\mathcal{R}}_{\phi\phi }+\bar{\mathcal{R}}_{zz}\right)}{ |\omega_{bd}|^2} +e^{-\frac{2 \sqrt{3} |\omega_{bd}|}{a}}\times\nonumber\\
&\left[\frac{\sqrt{3} a \left(4 \bar{\mathcal{R}}_{\rho \rho}+3 \bar{\mathcal{R}}_{\phi\phi }+\bar{\mathcal{R}}_{zz}\right)}{8 |\omega_{bd}|}+\frac{a^2 \left(26 \bar{\mathcal{R}}_{\rho \rho}+23 \bar{\mathcal{R}}_{\phi\phi }+5 \bar{\mathcal{R}}_{zz}\right)}{16 |\omega_{bd}|^2}+\frac{a^3 \left(58 \bar{\mathcal{R}}_{\rho \rho}+55 \bar{\mathcal{R}}_{\phi\phi }+13 \bar{\mathcal{R}}_{zz}\right)}{32 \sqrt{3} |\omega_{bd}|^3}\right]\nonumber\\
&+\frac{3 a L \left[\bar{\mathcal{R}}_{\rho \rho} \left(2p_{\rho }+q_{\rho }\right)-\bar{\mathcal{R}}_{zz} \left(2p_z+q_z\right)+2\sqrt{3} \bar{\mathcal{R}}_{z \rho } \left(q_{z\rho }-2p_{z\rho }\right)\right]+3\sqrt{3} \bar{\mathcal{R}}_{\phi\phi } \left(2p_{\phi }-q_{\phi }\right)}{64 a L^3 |\omega _{bd}|^2}\;\bigg \}\nonumber\\
&+\frac{ e^2}{3\pi}\sum_{\omega_b<\omega_d} \omega_{bd}^4
\, \bigg \{ e^{-\frac{2 \sqrt{3} |\omega_{bd}|}{a}}\left[\frac{\sqrt{3} a \left(4 \bar{\mathcal{R}}_{\rho \rho}+3 \bar{\mathcal{R}}_{\phi\phi }+\bar{\mathcal{R}}_{zz}\right)}{8 |\omega_{bd}|}+\frac{a^2 \left(26 \bar{\mathcal{R}}_{\rho \rho}+23 \bar{\mathcal{R}}_{\phi\phi }+5 \bar{\mathcal{R}}_{zz}\right)}{16 |\omega_{bd}|^2}\right.\nonumber\\
&\left.+\frac{a^3 \left(58 \bar{\mathcal{R}}_{\rho \rho}+55 \bar{\mathcal{R}}_{\phi\phi }+13 \bar{\mathcal{R}}_{zz}\right)}{32 \sqrt{3} |\omega_{bd}|^3}\right]+\frac{3 a L \left(\bar{\mathcal{R}}_{\rho \rho} q_{\rho }-\bar{\mathcal{R}}_{zz} q_z+2\sqrt{3} \bar{\mathcal{R}}_{z \rho } q_{z\rho }\right)-3\sqrt{3} \bar{\mathcal{R}}_{\phi\phi }q_{\phi }}{64 a L^3 |\omega _{bd}|^2}\;\bigg \}\;.\label{Lfartot}
\end{align}
In the equations above, 
\bea
p_{z}&=&p_{\rho}=\frac{2\sqrt{\sqrt{3}aL}|\omega_{bd}|}{a}\sin \frac{2\sqrt{\sqrt{3}aL}|\omega_{bd}|}{a}+9 \cos \frac{2\sqrt{\sqrt{3}aL}|\omega_{bd}|}{a} , \\
p_{\phi}&=& \frac{2\sqrt{\sqrt{3}aL}|\omega_{bd}|}{a}\sin \frac{2\sqrt{\sqrt{3}aL}|\omega_{bd}|}{a}- \cos \frac{2\sqrt{\sqrt{3}aL}|\omega_{bd}|}{a}, \\
p_{z\rho}&=& \frac{2\sqrt{\sqrt{3}aL}|\omega_{bd}|}{a}\sin \frac{2\sqrt{\sqrt{3}aL}|\omega_{bd}|}{a}+ \cos \frac{2\sqrt{\sqrt{3}aL}|\omega_{bd}|}{a}~~~~~~~~~~~
\eea
\end{widetext}
are oscillatory terms with respect to $L$, and 
\bea
q_{z}&=&q_{\rho}=\left(\frac{2\sqrt{\sqrt{3}aL}|\omega_{bd}|}{a}+9\right)e^{-\frac{2\sqrt{\sqrt{3}aL}|\omega_{bd}|}{a}},\\
q_{\phi}&=&\left(\frac{2\sqrt{\sqrt{3}aL}|\omega_{bd}|}{a}-1 \right)e^{-\frac{2\sqrt{\sqrt{3}aL}|\omega_{bd}|}{a}},\\
q_{z\rho}&=&\left(\frac{2\sqrt{\sqrt{3}aL}|\omega_{bd}|}{a}+1 \right)e^{-\frac{2\sqrt{\sqrt{3}aL}|\omega_{bd}|}{a}}~~~~~~~~~
\eea 
are exponential decay terms with respect to $L$. We can draw the following conclusions from the results above.

\emph{1.} In this regime, the leading contributions from vacuum fluctuations and radiation reaction are identical. However, the corrections arising from pure acceleration and the combined effect of acceleration and the boundary are different. Therefore, for ground-state atoms, spontaneous excitation is still possible.  

\emph{2.} 
By comparing Eqs.~\eqref{Lfarvf} and \eqref{Lfarrr} with Eqs.~\eqref{in-vf} and \eqref{in-rr}, it is evident that in this regime, the leading contribution comes from the inertial term in free space, while the effects of centripetal acceleration emerge only as higher-order corrections. 

\emph{3.} Comparing the results in this region with the free space case in the small acceleration limit (i.e., $a/\omega_{bd} \ll 1$), as given by Eqs. \eqref{free-space-vf} and \eqref{free-space-rr}, we find that the presence of the boundary introduces a subtle correction jointly governed by both the centripetal acceleration and the atom-boundary separation. This correction appears through the oscillatory terms ($p_{z}$, $p_{\phi}$, $p_{\rho}$) and  exponential decay terms ($q_{z}$, $q_{\phi}$, $q_{\rho}$), all of which depend on both the atom-boundary distance $L$ and the centripetal acceleration $a$. Notably, the rate of change of the mean atomic energy exhibits oscillations that scale as $\frac{2\sqrt{\sqrt{3}aL}|\omega_{bd}|}{a}$, in contrast to the inertial case, where the oscillation scales as $2 |\omega_{bd}| L$.  This oscillatory damping behavior with increasing atom-boundary distance contrasts sharply with the monotonic decay observed in the far region with $ 1/a\ll1/|\omega_{bd}| \ll L$.  
In particular, the contribution of vacuum fluctuations contains both oscillatory terms and exponential decay terms as functions of the distance $L$, whereas the contribution of radiation reaction contains only oscillatory terms.
When summing these contributions, distinct behavior emerges depending on the atomic state. For the ground state atoms  ($\omega_b < \omega_d$), the oscillatory terms ($p_{z}$, $p_{\phi}$, $p_{\rho}$, $p_{z\rho}$) associated with boundary corrections cancel out,  leaving only the exponential decay terms to influence the rate of change of the mean atomic energy. In contrast, for excited state atoms ($\omega_b > \omega_d$),  the oscillatory terms double in amplitude, amplifying the boundary effects.

At first glance, one might expect the boundary-related correction in the far regime to contribute only a higher-order correction to the free-space result. However, when the condition  $ \sqrt[5]{L^3/|\omega_{bd}|^2} \ll1/a \ll L$ is satisfied, the boundary term proportional to $ \frac{\bar{\mathcal{R}}_{z\rho}}{\sqrt{aL^3}|\omega_{bd}|}  \sin \frac{2\sqrt{\sqrt{3}aL}|\omega_{bd}|}{a} $ becomes significantly larger than the subleading free-space term proportional to $a^2/|\omega_{bd}|^2$. This reveals that the combined effect of acceleration and the presence of the boundary  can substantially alter the radiative properties of centripetally accelerated atoms, even at large atom-boundary separations.

\emph{4.} 
From Eqs.~\eqref{Lfarvf}$\sim$\eqref{Lfartot} and Eqs.~\eqref{free-space-vf}$\sim$\eqref{free-space-tot}, it is evident that both the rate of change of the mean atomic energy in free space and the boundary-induced corrections depend significantly on the atomic polarization direction. 
For the free-space term, the contributions from different atomic polarizations are of the same order although not identical.  In contrast, for the boundary-induced corrections, 
the contribution from polarization along the $\phi$-direction is significantly  smaller than those from other directions ($z$- or $\rho$-direction), due to the presence of an additional factor $1/aL$. This disparity  highlights the anisotropic nature of boundary-induced corrections, influenced by the interplay between centripetal acceleration and atomic polarization.

\subsection{Intermediate region between $1/a$ and $1/|\omega_{bd}|$}
Now, we consider the intermediate regime, in which the distance between the atom and the boundary $L$ is much smaller than either the characteristic length determined by the acceleration $1/a$ or the transition wavelength of the atom $1/|\omega_{bd}|$. This includes $1/a \ll L \ll 1/|\omega_{bd}|$ and $1/|\omega_{bd}| \ll L \ll 1/a$.

~

\subsubsection{Intermediate region with $1/a \ll L \ll 1/|\omega_{bd}|$ }
When the distance between the atom and the boundary $L$ is much larger than the characteristic length determined by the acceleration $1/a$, while it is much smaller than the transition wavelength of the atom $1/|\omega_{bd}|$, i.e.,  $1/a \ll L \ll 1/|\omega_{bd}|$, the contributions of vacuum fluctuations and radiation reaction to the rate of change of the mean atomic energy can be expressed as 
\begin{widetext}
  \begin{align}
{\biggl\langle {\frac{dH_A(\tau)}{d\tau}}\biggr\rangle}_{\rm VF} \approx
&{-\frac{ e^2}{3\pi}
} \left(\sum_{\omega_b>\omega_d}-\sum_{\omega_b<\omega_d}\right)\omega_{bd}^4
\,\bigg [\frac{a^3 \left(58 \bar{\mathcal{R}}_{\rho \rho}+55 \bar{\mathcal{R}}_{\phi\phi }+13 \bar{\mathcal{R}}_{zz}\right)}{32 \sqrt{3} |\omega _{bd}|^3}+\frac{\sqrt{3} a \left(15 \bar{\mathcal{R}}_{\phi\phi }+14 \bar{\mathcal{R}}_{\rho \rho}+5 \bar{\mathcal{R}}_{zz}\right)}{16 |\omega _{bd}|}\nonumber\\
&-\frac{9 \sqrt[4]{3} \sqrt{a L} \left(2 \bar{\mathcal{R}}_{z \rho }-\sqrt{3} \bar{\mathcal{R}}_{\rho \rho}+\sqrt{3} \bar{\mathcal{R}}_{zz}\right)}{128 L^3 |\omega _{bd}|^3}  
\;\bigg ]\;\label{Lnear1vf}
\end{align}
and
     \begin{align}
{\biggl\langle {\frac{dH_A(\tau)}{d\tau}}\biggr\rangle}_{\rm RR} \approx
&{-\frac{ e^2}{3\pi}
} \left(\sum_{\omega_b>\omega_d}+\sum_{\omega_b<\omega_d}\right)\omega_{bd}^4
\,\bigg [\frac{a^2 \left(4 \bar{\mathcal{R}}_{\rho \rho}+4\bar{\mathcal{R}}_{\phi\phi }+\bar{\mathcal{R}}_{zz}\right)}{2 | \omega _{bd}|^2}-\frac{3 \sqrt{3} \bar{\mathcal{R}}_{z \rho }}{8 L^2 | \omega _{bd}|^2}+\frac{\bar{\mathcal{R}}_{\rho \rho}+\bar{\mathcal{R}}_{\phi\phi }+\bar{\mathcal{R}}_{zz}}{2}\nonumber\\
&+\frac{3 \sqrt{3} \left(5 \bar{\mathcal{R}}_{\rho \rho}+2 \bar{\mathcal{R}}_{\phi\phi }-2 \bar{\mathcal{R}}_{zz}\right)}{16 a L^3 | \omega _{bd}|^2}\; \bigg ]\;,\label{Lnear1rr}
\end{align}
respectively. Then, the total rate of change of the mean atomic energy is obtained as
\begin{align}
{\biggl\langle {\frac{dH_A(\tau)}{d\tau}}\biggr\rangle}_{\rm tot} \approx
&{-\frac{ e^2}{3\pi}
} \sum_{\omega_b>\omega_d}\omega_{bd}^4
\,\bigg [\frac{a^3 \left(58 \bar{\mathcal{R}}_{\rho \rho}+55 \bar{\mathcal{R}}_{\phi\phi }+13 \bar{\mathcal{R}}_{zz}\right)}{32 \sqrt{3} |\omega _{bd}|^3}+\frac{a^2  \left(4 \bar{\mathcal{R}}_{\rho \rho}+4\bar{\mathcal{R}}_{\phi\phi }+\bar{\mathcal{R}}_{zz}\right)}{2 | \omega _{bd}|^2}\nonumber\\
&-\frac{9 \sqrt[4]{3} \sqrt{a L} \left(2 \bar{\mathcal{R}}_{z \rho }-\sqrt{3} \bar{\mathcal{R}}_{\rho \rho}+\sqrt{3} \bar{\mathcal{R}}_{zz}\right)}{128 L^3 |\omega _{bd}|^3}-\frac{3 \sqrt{3} \bar{\mathcal{R}}_{z \rho }}{8 L^2 | \omega _{bd}|^2}+\frac{\sqrt{3} a \left(15 \bar{\mathcal{R}}_{\phi\phi }+14 \bar{\mathcal{R}}_{\rho \rho}+5 \bar{\mathcal{R}}_{zz}\right)}{16 |\omega _{bd}|}\nonumber\\
&+\frac{\bar{\mathcal{R}}_{\rho \rho}+\bar{\mathcal{R}}_{\phi\phi }+\bar{\mathcal{R}}_{zz}}{2}+\frac{3 \sqrt{3} \left(5 \bar{\mathcal{R}}_{\rho \rho}+2 \bar{\mathcal{R}}_{\phi\phi }-2 \bar{\mathcal{R}}_{zz}\right)}{16 a L^3 | \omega _{bd}|^2}
\;\bigg ]\nonumber\\
&+\frac{ e^2}{3\pi}\sum_{\omega_b<\omega_d} \omega_{bd}^4
\, \bigg [ \frac{a^3 \left(58 \bar{\mathcal{R}}_{\rho \rho}+55 \bar{\mathcal{R}}_{\phi\phi }+13 \bar{\mathcal{R}}_{zz}\right)}{32 \sqrt{3} |\omega _{bd}|^3}-\frac{a^2  \left(4 \bar{\mathcal{R}}_{\rho \rho}+4\bar{\mathcal{R}}_{\phi\phi }+\bar{\mathcal{R}}_{zz}\right)}{2 | \omega _{bd}|^2}\nonumber\\
&-\frac{9 \sqrt[4]{3} \sqrt{a L} \left(2 \bar{\mathcal{R}}_{z \rho }-\sqrt{3} \bar{\mathcal{R}}_{\rho \rho}+\sqrt{3} \bar{\mathcal{R}}_{zz}\right)}{128 L^3 |\omega _{bd}|^3}+\frac{3 \sqrt{3} \bar{\mathcal{R}}_{z \rho }}{8 L^2 | \omega _{bd}|^2}+\frac{\sqrt{3} a \left(15 \bar{\mathcal{R}}_{\phi\phi }+14 \bar{\mathcal{R}}_{\rho \rho}+5 \bar{\mathcal{R}}_{zz}\right)}{16 |\omega _{bd}|}\nonumber\\
&-\frac{\bar{\mathcal{R}}_{\rho \rho}+\bar{\mathcal{R}}_{\phi\phi }+\bar{\mathcal{R}}_{zz}}{2}
-\frac{3 \sqrt{3} \left(5 \bar{\mathcal{R}}_{\rho \rho}+2 \bar{\mathcal{R}}_{\phi\phi }-2 \bar{\mathcal{R}}_{zz}\right)}{16 a L^3 | \omega _{bd}|^2}
\;\bigg ]\;.\label{Lnear1tot}
\end{align}
\end{widetext}

\emph{1.} In this region, the leading  contribution of vacuum fluctuations to the rate of change of the mean atomic energy scales as $a^3/|\omega_{bd}|^3$, which is significantly larger than the contribution from radiation reaction,  which scales as $a^2/|\omega_{bd}|^2$, given that  here $a/|\omega_{bd}| \gg 1$. Consequently, the contribution of vacuum fluctuations dominates the rate of change of the mean atomic energy for   both the ground- and excited-state atoms.

\emph{2.} 
In this region ($1/a \ll L \ll 1/|\omega_{bd}|$), the contribution of vacuum fluctuations 
retains the same functional form as that in the far region with $1/a \ll 1/|\omega_{bd}| \ll L$. This sharply contrasts with the inertial case, where the behavior differs significantly between the near-region $L\ll 1/|\omega_{bd}|$ and the far-region $L\gg 1/|\omega_{bd}|$, indicating that large centripetal acceleration profoundly modifies atomic radiative behavior. In order to further highlight the role of acceleration, we present below the contributions of vacuum fluctuations and radiation reaction to the rate of change of the mean atomic energy for inertial atoms when the atom-boundary distance $L$ is much smaller than the transition wavelength of the atoms $1/|\omega_{bd}|$:
\begin{widetext}
\begin{align}
{\biggl\langle {\frac{dH_A(\tau)}{d\tau}}\biggr\rangle}_{\rm VF}^{\rm in} \approx
&{-\frac{ e^2}{3\pi}
} \left(\sum_{\omega_b>\omega_d}-\sum_{\omega_b<\omega_d}\right)\omega_{bd}^4
\,\bigg [\bar{\mathcal{R}}_{zz}+\frac{1}{5} L^2 \omega_{bd}^2 \left(2 \bar{\mathcal{R}}_{\rho \rho}+2 \bar{\mathcal{R}}_{\phi\phi }-\bar{\mathcal{R}}_{zz}\right)\;\bigg ]\;\label{invfLsmall}
\end{align}
and
\begin{align}
{\biggl\langle {\frac{dH_A(\tau)}{d\tau}}\biggr\rangle}_{\rm RR}^{\rm in} \approx
&{-\frac{ e^2}{3\pi}
} \left(\sum_{\omega_b>\omega_d}+\sum_{\omega_b<\omega_d}\right)\omega_{bd}^4
\,\bigg [\bar{\mathcal{R}}_{zz}+\frac{1}{5} L^2 \omega_{bd}^2 \left(2 \bar{\mathcal{R}}_{\rho \rho}+2 \bar{\mathcal{R}}_{\phi\phi }-\bar{\mathcal{R}}_{zz}\right)\;\bigg ]\;,\label{inrrLsmall}
\end{align}
\end{widetext}
respectively. Comparing Eqs.~\eqref{invfLsmall} and \eqref{inrrLsmall} with Eqs.~\eqref{Lnear1vf} and \eqref{Lnear1rr}, we find that, in the inertial case, the dominant contribution arises from atoms polarizable along the direction normal to the boundary. In contrast, in the presence of centripetal acceleration, contributions from all polarization directions are of the same order of magnitude. 

\emph{3.} Furthermore, comparing Eqs.~(\ref{Lnear1vf}) and (\ref{Lnear1rr})  with their free-space counterparts in Eqs.~(\ref{freevf}) and (\ref{freerr}), we find that  the free-space result can be recovered by neglecting all $L$-dependent terms in the presence of a reflecting boundary.  
Notably, the dominant boundary-induced correction to the radiation reaction term, scaling as $\frac{1}{L^2|\omega_{bd}|^2}$, originates from a $z\rho$ cross-polarization component, a feature absent in the inertial case. Although this correction does not explicitly depend on the centripetal acceleration $a$, it arises from the combined effect of the boundary and acceleration. 
Moreover, 
when the condition $1/a \ll L\ll1/\sqrt[5]{a^3|\omega_{bd}|^2}$ is satisfied, the correction due to the combined effects of acceleration and the boundary, proportional to $\frac{\sqrt{aL}}{L^3|\omega_{bd}|^3}$, becomes significantly larger than the subleading free-space term proportional to $\frac{a^2}{|\omega_{bd}|^2}$.  This reveals that, even in the near-intermediate region, where the atom is far from the acceleration scale but not yet beyond the transition wavelength, boundary effects enhanced by acceleration can strongly influence the atom's radiative properties.

\emph{4.} In this region, the contributions of transverse polarization (atomic polarization perpendicular to the rotation axis) and axial polarization (atomic polarization parallel to the rotation axis) to the rate of change of the mean energy of the atom are of the same order of magnitude for both ground-state and excited-state atoms, as shown in Eq.~(\ref{Lnear1tot}). In addition, for the distance-dependent terms, 
the contribution from polarization along the $\phi$-direction (proportional to $\frac{1}{aL^3|\omega_{bd}|^2}$) is always much smaller than that from other atomic polarizations (proportional to $\frac{\sqrt{a L}}{L^3 |\omega _{bd}|^3}$), since their ratio $\frac{|\omega_{bd}|}{a\sqrt{a L}}$ is much less than 1. 

~

\subsubsection{Intermediate region with $1/|\omega_{bd}| \ll L \ll 1/a$ }
When the distance $L$ between the atom and the boundary is much smaller than the characteristic length determined by the acceleration $1/a$, while much larger than the characteristic length determined by the transition wavelength of the atom $1/|\omega_{bd}|$, i.e., $1/|\omega_{bd}| \ll L \ll 1/a$, the contributions of vacuum fluctuations and radiation reaction to the rate of change of the mean atomic energy can be expressed as
\begin{widetext}
\begin{align}
{\biggl\langle {\frac{dH_A(\tau)}{d\tau}}\biggr\rangle}_{\rm VF} \approx
&{-\frac{ e^2}{3\pi}
} \left(\sum_{\omega_b>\omega_d}-\sum_{\omega_b<\omega_d} \right)\omega_{bd}^4
\,\bigg \{\frac{\bar{\mathcal{R}}_{\rho \rho}+\bar{\mathcal{R}}_{\phi\phi }+\bar{\mathcal{R}}_{zz}}{2} +\frac{\sin \left(2 L | \omega _{bd}|\right)}{32 L^3 | \omega _{bd}|^3} \left[-3 \left(4 L^2 |\omega _{bd}|^2-1\right) \left(\bar{\mathcal{R}}_{\rho \rho}+\bar{\mathcal{R}}_{\phi \phi}\right)\right.\nonumber\\
&\left.+6 \bar{\mathcal{R}}_{zz}+24 a L^3 |\omega _{bd}|^2 \bar{\mathcal{R}}_{z\rho }+4 a^2 L^4 |\omega _{bd}|^2 \left(10 \bar{\mathcal{R}}_{\rho \rho}+7 \bar{\mathcal{R}}_{\phi \phi}-4 \bar{\mathcal{R}}_{zz}\right)
\right]+\frac{\cos \left(2 L | \omega _{bd}|\right)}{16 L^2 | \omega _{bd}|^2} \left[\left(\bar{\mathcal{R}}_{\rho \rho}+\bar{\mathcal{R}}_{\phi\phi }\right)\times \right.\nonumber\\
&\left.\left(2 a^2 L^4 | \omega _{bd}|^2-3\right)-6 \bar{\mathcal{R}}_{zz}-2 a L \bar{\mathcal{R}}_{z \rho } \left(2 a^2 L^4 | \omega _{bd}|^2+3\right)\right] +\frac{a^2 \left[4 \left(\bar{\mathcal{R}}_{\rho \rho}+\bar{\mathcal{R}}_{\phi\phi }\right)+\bar{\mathcal{R}}_{zz}\right]}{2 | \omega _{bd}|^2}\nonumber\\
&+\frac{\sqrt{3} a e^{-\frac{2 \sqrt{3} |\omega_{bd}|}{a}}}{8 |\omega_{bd}|} \left[4 \bar{\mathcal{R}}_{\rho \rho}+3 \bar{\mathcal{R}}_{\phi\phi }+\bar{\mathcal{R}}_{zz}-e^{-\frac{a L^2 |\omega _{bd}|}{\sqrt{3}}}\left(4 \bar{\mathcal{R}}_{\rho \rho}+3 \bar{\mathcal{R}}_{\phi\phi }-\bar{\mathcal{R}}_{zz}-6aL\bar{\mathcal{R}}_{z\rho}\right)\right]\;\bigg \} \;\label{Lnear3vf}
\end{align}
and
\begin{align}
{\biggl\langle {\frac{dH_A(\tau)}{d\tau}}\biggr\rangle}_{\rm RR} \approx
&{-\frac{ e^2}{3\pi}
}\left(\sum_{\omega_b>\omega_d}+\sum_{\omega_b<\omega_d} \right)\omega_{bd}^4
\,\bigg \{\frac{\bar{\mathcal{R}}_{\rho \rho}+\bar{\mathcal{R}}_{\phi\phi }+\bar{\mathcal{R}}_{zz}}{2} +\frac{\sin \left(2 L | \omega _{bd}|\right)}{32 L^3 | \omega _{bd}|^3} \left[-3 \left(4 L^2 |\omega _{bd}|^2-1\right) \left(\bar{\mathcal{R}}_{\rho \rho}+\bar{\mathcal{R}}_{\phi \phi}\right)\right.\nonumber\\
&\left.+6 \bar{\mathcal{R}}_{zz}+24 a L^3 |\omega _{bd}|^2 \bar{\mathcal{R}}_{z\rho }+4 a^2 L^4 |\omega _{bd}|^2 \left(10 \bar{\mathcal{R}}_{\rho \rho}+7 \bar{\mathcal{R}}_{\phi \phi}-4 \bar{\mathcal{R}}_{zz}\right)
\right]+\frac{\cos \left(2 L | \omega _{bd}|\right)}{16 L^2 | \omega _{bd}|^2} \left[\left(\bar{\mathcal{R}}_{\rho \rho}+\bar{\mathcal{R}}_{\phi\phi }\right)\times\right.\nonumber\\
&\left. \left(2 a^2 L^4 | \omega _{bd}|^2-3\right)-6 \bar{\mathcal{R}}_{zz}-2 a L \bar{\mathcal{R}}_{z \rho } \left(2 a^2 L^4 | \omega _{bd}|^2+3\right)\right] +\frac{a^2 \left[4 \left(\bar{\mathcal{R}}_{\rho \rho}+\bar{\mathcal{R}}_{\phi\phi }\right)+\bar{\mathcal{R}}_{zz}\right]}{2 | \omega _{bd}|^2}\;\bigg \} \;,\label{Lnear3rr}
\end{align}
respectively. Adding the equations above yields the total rate of change of the mean atomic energy
\begin{align}
{\biggl\langle {\frac{dH_A(\tau)}{d\tau}}\biggr\rangle}_{\rm tot} \approx
&{-\frac{ e^2}{3\pi}
}\sum_{\omega_b>\omega_d}\omega_{bd}^4
\,\bigg \{\bar{\mathcal{R}}_{\rho \rho}+\bar{\mathcal{R}}_{\phi\phi }+\bar{\mathcal{R}}_{zz}  +\frac{\sin \left(2 L | \omega _{bd}|\right)}{16 L^3 | \omega _{bd}|^3} \left[-3 \left(4 L^2 |\omega _{bd}|^2-1\right) \left(\bar{\mathcal{R}}_{\rho \rho}+\bar{\mathcal{R}}_{\phi \phi}\right)+6 \bar{\mathcal{R}}_{zz}\right.\nonumber\\
&\left.+24 a L^3 |\omega _{bd}|^2 \bar{\mathcal{R}}_{z\rho }+4 a^2 L^4 |\omega _{bd}|^2 \left(10 \bar{\mathcal{R}}_{\rho \rho}+7 \bar{\mathcal{R}}_{\phi \phi}-4 \bar{\mathcal{R}}_{zz}\right)
\right]+\frac{\cos \left(2 L | \omega _{bd}|\right)}{8 L^2 | \omega _{bd}|^2} \left[\left(\bar{\mathcal{R}}_{\rho \rho}+\bar{\mathcal{R}}_{\phi\phi }\right)\times\right.\nonumber\\
&\left. \left(2 a^2 L^4 | \omega _{bd}|^2-3\right)-6 \bar{\mathcal{R}}_{zz}-2 a L \bar{\mathcal{R}}_{z \rho } \left(2 a^2 L^4 | \omega _{bd}|^2+3\right)\right]+\frac{a^2 \left[4 \left(\bar{\mathcal{R}}_{\rho \rho}+\bar{\mathcal{R}}_{\phi\phi }\right)+\bar{\mathcal{R}}_{zz}\right]}{ | \omega _{bd}|^2}\nonumber\\
&+ \frac{\sqrt{3} a e^{-\frac{2 \sqrt{3} |\omega_{bd}|}{a}}}{8 |\omega_{bd}|} \left[4 \bar{\mathcal{R}}_{\rho \rho}+3 \bar{\mathcal{R}}_{\phi\phi }+\bar{\mathcal{R}}_{zz}-e^{-\frac{a L^2 |\omega _{bd}|}{\sqrt{3}}}\left(4 \bar{\mathcal{R}}_{\rho \rho}+3 \bar{\mathcal{R}}_{\phi\phi }-\bar{\mathcal{R}}_{zz}-6aL\bar{\mathcal{R}}_{z\rho}\right)\right]\;\bigg \}\nonumber\\
&+\frac{ e^2}{3\pi}\sum_{\omega_b<\omega_d} \omega_{bd}^4
\, \frac{\sqrt{3} a e^{-\frac{2 \sqrt{3} |\omega_{bd}|}{a}}}{8 |\omega_{bd}|} \left[4 \bar{\mathcal{R}}_{\rho \rho}+3 \bar{\mathcal{R}}_{\phi\phi }+\bar{\mathcal{R}}_{zz}-e^{-\frac{a L^2 |\omega _{bd}|}{\sqrt{3}}}\left(4 \bar{\mathcal{R}}_{\rho \rho}+3 \bar{\mathcal{R}}_{\phi\phi }-\bar{\mathcal{R}}_{zz}-6aL\bar{\mathcal{R}}_{z\rho}\right)\right]\;.\label{Lnear3tot}
\end{align}
\end{widetext}

\emph{1.} In this region, as shown in Eq. (\ref{Lnear3tot}), the rate of change of the mean atomic energy for atoms in the ground state ($\omega_b < \omega_d$) is extremely small. This is primarily due to the presence  of an exponential suppression factor  $e^{-\frac{2 \sqrt{3} | \omega _{bd}|}{a}}$, with $|\omega _{bd}| / a \gg 1$.  The suppression arises because the contributions of vacuum fluctuations and radiation reaction to the rate of change of the mean atomic energy are nearly equal  in magnitude but opposite in sign. Consequently they almost cancel each other out, leaving only terms with the exponential suppression factor $e^{-\frac{2 \sqrt{3} | \omega _{bd}|}{a}}$, as shown in Eqs. \eqref{Lnear3vf} and \eqref{Lnear3rr}.  This behavior aligns with physical intuition, because the characteristic energy determined by the acceleration is much smaller than the energy level spacing of the atom. As a result, the atom lacks sufficient energy to undergo spontaneous excitation, making such events highly improbable.

\emph{2.} A comparison of Eqs.~\eqref{Lnear3vf} and \eqref{Lnear3rr} with Eqs.~(\ref{free-space-vf}) and (\ref{free-space-rr}) reveals that, 
if all the distance-dependent terms in Eqs.~\eqref{Lnear3vf} and \eqref{Lnear3rr} are neglected, the results reduce to those in the free-space case, as given by Eqs.~(\ref{free-space-vf}) and (\ref{free-space-rr}). 
This indicates that all boundary-induced effects are included in the terms that depend on the atom-boundary distance $L$. 
For atoms in an excited state ($\omega_b>\omega_d$), 
the boundary-induced correction to the rate of change of the mean atomic energy depends solely on the boundary, in contrast to previously discussed cases where the leading correction arises from the combined effect of the boundary and centripetal acceleration. Although this correction remains much smaller than the leading free-space term, it exceeds the subleading one. 
However, for atoms in the ground state ($\omega_b<\omega_d$),  the situation is markedly different. 
In this region ($1/|\omega_{bd}| \ll L \ll 1/a$), 
the boundary-related correction comes from the combined effect of the boundary and centripetal acceleration. 
The ratio of the boundary-induced correction term for ground-state atoms to the free-space term is $e^{-\frac{a L^2 |\omega _{bd}|}{\sqrt{3}}}$, making it unclear which term dominates. By  further subdividing this region, it becomes evident that when $ 1/\sqrt{a|\omega_{bd}|} \ll L \ll 1/a$, the boundary-induced correction is negligible, showing an extremely weak effect of the boundary on the spontaneous excitation of ground-state atoms. Conversely, when $L$ lies in the range $1/|\omega_{bd}| \ll L \ll 1/\sqrt{a|\omega_{bd}|}$, the boundary-induced correction can become comparable to the free-space result, indicating a substantial modification of the radiative properties of the atom due to the boundary.

\emph{3.} 
In this region, the contributions from atomic polarization along the $\rho$-, $\phi$-, and $z$-directions to the boundary-induced corrections in the rate of change of the mean atomic energy for ground-state atoms are of the same order, and significantly larger than that from the $z\rho$ cross term, as the ratio $a L\ll 1$. However, for atoms in an excited state, the contributions from atomic polarization along the tangential directions ($\rho$- or $\phi$-direction) are much larger than that from the normal direction ($z$-direction), since $L^2 |\omega_{bd}|^2 \gg1$.

\subsection{Near regions with $L \ll 1/a$ and  $L \ll1/|\omega_{bd}|$}
We now consider the scenario when the distance between the atom and the boundary $L$ is much smaller than both the characteristic length determined by the acceleration $1/a$ and the transition wavelength of the atom $1/|\omega_{bd}|$. This defines the near region, which encompasses two distinct cases: $L \ll 1/a \ll 1/|\omega_{bd}|$ and $L \ll 1/|\omega_{bd}| \ll 1/a$.

~

\subsubsection{ Near region with $L \ll 1/a \ll 1/|\omega_{bd}|$ }
When the distance between the atom and the boundary $L$ is much smaller than the characteristic length determined by the acceleration $1/a$, and $1/a$ is significantly smaller than the transition wavelength of the atom $1/|\omega_{bd}|$, i.e., $L \ll 1/a \ll 1/|\omega_{bd}|$, the contributions of vacuum fluctuations and radiation reaction to the rate of change of the mean atomic energy are given by
\begin{widetext}
\begin{align}
{\biggl\langle {\frac{dH_A(\tau)}{d\tau}}\biggr\rangle}_{\rm VF} \approx
&{-\frac{ e^2}{3\pi}
} \left(\sum_{\omega_b>\omega_d}-\sum_{\omega_b<\omega_d} \right)\omega_{bd}^4
\,\bigg \{\bar{\mathcal{R}}_{zz} \left(\frac{13 a^3}{16 \sqrt{3} | \omega _{bd}|^3}+\frac{5 \sqrt{3} a}{8 | \omega _{bd}|}\right)+a L \bar{\mathcal{R}}_{z \rho } \left(\frac{65 a^3}{16 \sqrt{3} | \omega _{bd}|^3}+\frac{21 \sqrt{3} a}{8 | \omega _{bd}|}\right)\nonumber\\
&+\frac{a^5  L^2\left( 504 \bar{\mathcal{R}}_{\rho \rho}+399 \bar{\mathcal{R}}_{\phi\phi }-147 \bar{\mathcal{R}}_{zz}\right)}{64 \sqrt{3} | \omega _{bd}|^3}\;\bigg \} \;\label{Lnear2vf}
\end{align}
and
\begin{align}
{\biggl\langle {\frac{dH_A(\tau)}{d\tau}}\biggr\rangle}_{\rm RR} \approx
&{-\frac{ e^2}{3\pi}
}\left(\sum_{\omega_b>\omega_d}+\sum_{\omega_b<\omega_d} \right)\omega_{bd}^4
\,\bigg [\bar{\mathcal{R}}_{zz} \left(\frac{a^2}{| \omega _{bd}|^2}+1\right)+a L \bar{\mathcal{R}}_{z \rho } \frac{ 5 a^2}{| \omega _{bd}|^2}+\frac{a^4 L^2 \left(10 \bar{\mathcal{R}}_{\rho \rho}-3 \bar{\mathcal{R}}_{zz}+8 \bar{\mathcal{R}}_{\phi\phi }\right)}{| \omega _{bd}|^2}\; \bigg ] \;,\label{Lnear2rr}
\end{align}
respectively. The total rate of change of the mean atomic energy is then derived by summing up Eqs. \eqref{Lnear2vf} and \eqref{Lnear2rr} as  
\begin{align}
{\biggl\langle {\frac{dH_A(\tau)}{d\tau}}\biggr\rangle}_{\rm tot}\approx
&{-\frac{ e^2}{3\pi}
} \sum_{\omega_b>\omega_d}\omega_{bd}^4
\,\bigg \{\bar{\mathcal{R}}_{zz} \left(\frac{13 a^3}{16 \sqrt{3} | \omega _{bd}|^3}+\frac{a^2}{| \omega _{bd}|^2}+\frac{5 \sqrt{3} a}{8 | \omega _{bd}|}+1\right)+a L \bar{\mathcal{R}}_{z \rho } \left(\frac{65 a^3}{16 \sqrt{3} | \omega _{bd}|^3}+ \frac{ 5 a^2}{ | \omega _{bd}|^2}\right.\nonumber\\
&\left.+\frac{21 \sqrt{3} a}{8 | \omega _{bd}|}\right)
+a^2 L^2\left[\frac{a^3 \left(504 \bar{\mathcal{R}}_{\rho \rho}+399 \bar{\mathcal{R}}_{\phi\phi }-147 \bar{\mathcal{R}}_{zz}\right)}{64 \sqrt{3} | \omega _{bd}|^3}+\frac{a^2 \left(10 \bar{\mathcal{R}}_{\rho \rho}+8 \bar{\mathcal{R}}_{\phi\phi }-3 \bar{\mathcal{R}}_{zz}\right)}{| \omega _{bd}|^2}\right]\;\bigg \}\nonumber\\
&+\frac{ e^2}{3\pi}\sum_{\omega_b<\omega_d} \omega_{bd}^4
\, \bigg \{ \bar{\mathcal{R}}_{zz} \left(\frac{13 a^3}{16 \sqrt{3} | \omega _{bd}|^3}-\frac{a^2}{| \omega _{bd}|^2}+\frac{5 \sqrt{3} a}{8 | \omega _{bd}|}-1\right)+a L \bar{\mathcal{R}}_{z \rho } \left(\frac{65 a^3}{16 \sqrt{3} | \omega _{bd}|^3}- \frac{ 5 a^2}{ | \omega _{bd}|^2}\right.\nonumber\\
&\left.+\frac{21 \sqrt{3} a}{8 | \omega _{bd}|}\right)
+a^2 L^2\left[\frac{a^3 \left(504\bar{\mathcal{R}}_{\rho \rho}+399 \bar{\mathcal{R}}_{\phi\phi }-147 \bar{\mathcal{R}}_{zz}\right)}{64 \sqrt{3} | \omega _{bd}|^3}-\frac{a^2 \left(10 \bar{\mathcal{R}}_{\rho \rho}+8 \bar{\mathcal{R}}_{\phi\phi }-3 \bar{\mathcal{R}}_{zz}\right)}{| \omega _{bd}|^2}\right]\;\bigg \} \;.\label{Lnear2tot}
\end{align}
\end{widetext}

\emph{1.} Similar to the result in the region $1/a \ll L \ll 1/|\omega_{bd}|$, the contribution of vacuum fluctuations to the rate of change of the mean atomic energy outweighs that of radiation reaction.     Specifically, the leading term of vacuum fluctuations  
scales as $a^3/|\omega_{bd}|^3$,  while the leading term of radiation reaction 
scales as   $a^2/|\omega_{bd}|^2$. Since $a/|\omega_{bd}| \gg 1$,  the vacuum fluctuation contribution is significantly larger. Consequently, the total rates of change of the mean atomic energy for atoms in both the ground state and the excited state are of the same order of magnitude, as shown in Eq.~(\ref{Lnear2tot}).

\emph{2.} Comparing Eqs. (\ref{Lnear2vf}) and (\ref{Lnear2rr}) with Eqs. (\ref{freevf}) and (\ref{freerr}) reveals that, in this region, the result for the free-space case cannot be recovered by simply neglecting the distance-dependent terms present in the case with a reflecting boundary. This shows that the boundary induces a significant correction comparable in magnitude to those in the free-space case. 
This is because the presence of the boundary causes a significant modification to the fluctuating electromagnetic fields near the boundary, as the fields must satisfy the boundary condition that the electric field on the surface of a perfectly conducting plane is normal to the plane.

\emph{3.} Correspondingly, in this region, the contributions from different polarization directions to the rate of change of the mean atomic energy vary significantly. 
The leading contributions for both ground-state and excited-state atoms come from the axial polarization (normal to the boundary), which are
proportional to $a^3/|\omega_{bd}|^3$. These contributions  are  twice as large as the corresponding terms in the free-space case [see Eqs.~(\ref{freevf}) and (\ref{freerr})].   The enhancement arises because the reflecting boundary doubles the normal component of the electric field. The contributions from the $z\rho$ cross term for both ground-state and excited-state atoms, proportional to $a^4 L/|\omega_{bd}|^3$,  are smaller than that of the axial polarization but still substantial.  These terms arise due to the interaction between the normal and tangential components of the electric field near the boundary.  
The contributions from transverse polarization are the smallest, being proportional to  $a^5 L^2/|\omega_{bd}|^3$ for both ground-state and excited-state atoms. This suppression can be attributed to the boundary conditions, as the tangential components of the electric field vanish at the reflecting boundary, significantly reducing their impact on the atom's radiative properties.

\subsubsection{ Near region with $L \ll 1/|\omega_{bd}| \ll 1/a$}

When the distance between the atom and the boundary $L$ is much smaller than the transition wavelength of the atom $1/|\omega_{bd}|$, and simultaneously $1/|\omega_{bd}|$ is much smaller than the characteristic length determined by the acceleration $1/a$, i.e., $L \ll 1/|\omega_{bd}| \ll 1/a$, the contributions of vacuum fluctuations and radiation reaction to the rate of change of the mean atomic energy can be expressed as
\begin{widetext}
\begin{align}
{\biggl\langle {\frac{dH_A(\tau)}{d\tau}}\biggr\rangle}_{\rm VF} \approx
&{-\frac{ e^2}{3\pi}
} \left(\sum_{\omega_b>\omega_d}-\sum_{\omega_b<\omega_d} \right)\omega_{bd}^4
\,\bigg \{ \bar{\mathcal{R}}_{zz} \left(1+\frac{a^2}{| \omega _{bd}|^2}\right)+2a L \bar{\mathcal{R}}_{z \rho } \left(1+\frac{5 a^2}{2 | \omega _{bd}|^2}\right)\nonumber\\
&+\frac{L^2 | \omega _{bd}|^2}{5}  \left(2 \bar{\mathcal{R}}_{\rho \rho}+2\bar{\mathcal{R}}_{\phi\phi }-\bar{\mathcal{R}}_{zz}\right)
+a^2 L^2 \left(6 \bar{\mathcal{R}}_{\rho \rho}+5 \bar{\mathcal{R}}_{\phi\phi }-2 \bar{\mathcal{R}}_{zz}\right)-\frac{3}{35} L^4 | \omega _{bd}|^4 \left(\bar{\mathcal{R}}_{\rho \rho}+\bar{\mathcal{R}}_{\phi\phi }\right)\nonumber\\
&+e^{-\frac{2 \sqrt{3} | \omega _{bd}|}{a}} \left[\frac{\sqrt{3} a \bar{\mathcal{R}}_{zz}}{4 | \omega _{bd}|}+\frac{3 \sqrt{3} a^2 L \bar{\mathcal{R}}_{z \rho }}{4 | \omega _{bd}|}+\frac{1}{8} a^2 L^2 \left(4 \bar{\mathcal{R}}_{\rho \rho}+3 \bar{\mathcal{R}}_{\phi\phi }\right)\right]\;\bigg \} \;\label{Lnear4vf}
\end{align}
and
\begin{align}
{\biggl\langle {\frac{dH_A(\tau)}{d\tau}}\biggr\rangle}_{\rm RR} \approx
&{-\frac{ e^2}{3\pi}
}\left(\sum_{\omega_b>\omega_d}+\sum_{\omega_b<\omega_d} \right)\omega_{bd}^4
\,\bigg \{  \bar{\mathcal{R}}_{zz} \left(1+\frac{a^2}{| \omega _{bd}|^2}\right)+2a L \bar{\mathcal{R}}_{z \rho } \left(1+\frac{5 a^2}{2 | \omega _{bd}|^2}\right)\nonumber\\
&+\frac{L^2 | \omega _{bd}|^2}{5}  \left(2 \bar{\mathcal{R}}_{\rho \rho}+2\bar{\mathcal{R}}_{\phi\phi }-\bar{\mathcal{R}}_{zz}\right)
+a^2 L^2 \left(6 \bar{\mathcal{R}}_{\rho \rho}+5 \bar{\mathcal{R}}_{\phi\phi }-2 \bar{\mathcal{R}}_{zz}\right)-\frac{3}{35} L^4 | \omega _{bd}|^4 \left(\bar{\mathcal{R}}_{\rho \rho}+\bar{\mathcal{R}}_{\phi\phi }\right)\; \bigg \} \;,\label{Lnear4rr}
\end{align}
respectively, and the total rate of change of the mean atomic energy can then be obtained as
\begin{align}
{\biggl\langle {\frac{dH_A(\tau)}{d\tau}}\biggr\rangle}_{\rm tot} \approx
&{-\frac{ e^2}{3\pi}
}\sum_{\omega_b>\omega_d}\omega_{bd}^4
\,\bigg \{ 2\bar{\mathcal{R}}_{zz} \left(1+\frac{a^2}{| \omega _{bd}|^2}\right)+4a L \bar{\mathcal{R}}_{z \rho } \left(1+\frac{5 a^2}{2 | \omega _{bd}|^2}\right)\nonumber\\
&+\frac{2L^2 | \omega _{bd}|^2}{5}  \left(2 \bar{\mathcal{R}}_{\rho \rho}+2\bar{\mathcal{R}}_{\phi\phi }-\bar{\mathcal{R}}_{zz}\right)
+2a^2 L^2 \left(6 \bar{\mathcal{R}}_{\rho \rho}+5 \bar{\mathcal{R}}_{\phi\phi }-2 \bar{\mathcal{R}}_{zz}\right)-\frac{6}{35} L^4 | \omega _{bd}|^4 \left(\bar{\mathcal{R}}_{\rho \rho}+\bar{\mathcal{R}}_{\phi\phi }\right)\nonumber\\
&+e^{-\frac{2 \sqrt{3} | \omega _{bd}|}{a}} \left(\frac{\sqrt{3} a \bar{\mathcal{R}}_{zz}}{4 | \omega _{bd}|}+\frac{3 \sqrt{3} a^2 L \bar{\mathcal{R}}_{z \rho }}{4 | \omega _{bd}|}+\frac{1}{8} a^2 L^2 \left(4 \bar{\mathcal{R}}_{\rho \rho}+3 \bar{\mathcal{R}}_{\phi\phi }\right)\right)\;\bigg \}\nonumber\\
&+\frac{ e^2}{3\pi}\sum_{\omega_b<\omega_d} \omega_{bd}^4
\, \bigg \{ e^{-\frac{2 \sqrt{3} | \omega _{bd}|}{a}} \left[\frac{\sqrt{3} a \bar{\mathcal{R}}_{zz}}{4 | \omega _{bd}|}+\frac{3 \sqrt{3} a^2 L \bar{\mathcal{R}}_{z \rho }}{4 | \omega _{bd}|}+\frac{1}{8} a^2 L^2 \left(4 \bar{\mathcal{R}}_{\rho \rho}+3 \bar{\mathcal{R}}_{\phi\phi }\right)\right]\;\bigg \} \;.\label{Lnear4tot}
\end{align}
\end{widetext}

\emph{1.} Similar to the case when $1/|\omega_{bd}| \ll L \ll 1/a$, the contributions of vacuum fluctuations and radiation reaction to the rate of change of the mean atomic energy in the ground state  are nearly equal in magnitude but opposite in sign, leading to their near cancellation. As a result, the net rate of change of the mean atomic energy becomes exceedingly small.  This near cancellation significantly suppresses the possibility of spontaneous excitation for a ground-state atom, as the energy required for excitation is much larger than the characteristic energy scale determined by the acceleration.  Consequently, spontaneous excitation of a ground-state atom is highly unlikely. This behavior is consistent across regions whether the distance $L$ is small or large compared to the transition wavelength $1/|\omega_{bd}|$.

\emph{2.} In this region, a comparison of Eqs.~(\ref{Lnear4vf}) and (\ref{Lnear4rr}) with Eqs.~(\ref{free-space-vf}) and (\ref{free-space-rr}) shows that, 
similar to the region $L \ll 1/a \ll 1/|\omega_{bd}|$, 
simply neglecting the distance-dependent terms does not yield the free-space result. 
This shows that, when the atom-boundary distance $L$ is much smaller than both  $1/a$ and $1/|\omega_{bd}|$, 
the presence of the reflecting boundary significantly affects the rate of change of the mean atomic energy. 
This effect arises from the interaction between the atom and the modified vacuum electromagnetic fields near the boundary, which must satisfy the boundary condition that the electric field on the surface of a perfectly conducting plane is normal to the plane.

\emph{3.} Correspondingly, similar to the region $L \ll 1/a \ll 1/|\omega_{bd}|$, the dominant contribution in this region 
comes from the axial polarization (normal to the boundary), which is independent of the atom-boundary distance $L$ and is twice that in the free-space case. 
 This can be observed by comparing  Eqs.~(\ref{Lnear4vf}) and (\ref{Lnear4rr}) with Eqs.~(\ref{free-space-vf}) and (\ref{free-space-rr}). 
Compared to the terms contributed by axial polarization,   the $z\rho$ cross terms for both ground-state and excited-state atoms contain an additional factor of $aL$,  which remain significant, although much smaller than the contribution from the axial polarization. 
For the contribution of the transverse polarization (tangential to the boundary), 
there is an additional factor $L^2\omega_{bd}^2$ for excited-state atoms, and $a^2L^2$ for ground-state atoms, compared to the contribution from axial polarization (tangential to the boundary). This makes 
its contribution significantly smaller than that of the axial polarization.  
This disparity stems from the fact that the tangential components of the electric field are suppressed (vanish) at the reflecting boundary,  while the normal component of the electric field is enhanced by the boundary, effectively doubling its magnitude.

\section{Summary}\label{sec5}
In this paper, we have investigated the rate of change of the mean atomic energy for centripetally accelerated atoms interacting with fluctuating electromagnetic fields in a vacuum near a reflecting boundary,  employing the DDC formalism. 
We performed a comprehensive analysis of the  individual  contributions from vacuum fluctuations and radiation reaction to this rate, focusing on the effects of the boundary and centripetal acceleration, both of which are closely dependent on the atomic polarization.

We find that the presence of a reflecting boundary significantly alters the radiative properties of centripetally accelerated atoms, particularly when the atom-boundary distance $L$ is much smaller than both the characteristic length determined by the acceleration ($1/a$) and the transition wavelength of the atom ($1/|\omega_{bd}|$).  In this near-boundary regime,  the boundary induces significant corrections to the rate of change of the mean atomic energy, which are comparable in magnitude to those in the free-space case.  Specifically,  the contribution from axial polarization (normal to the boundary) doubles compared to the free-space scenario due to the boundary-enhanced normal electric field component, while the transverse polarization contributions (tangential to the boundary) are nearly negligible, as the tangential electric field components vanish at the boundary. These effects demonstrate the boundary's strong influence on atomic radiative properties in close proximity. 

The role of centripetal acceleration depends sensitively on its magnitude relative to the atomic transition frequency.  When the centripetal acceleration is much larger than the characteristic acceleration determined by the transition frequency of the atom, the contribution from vacuum fluctuations dominates over that from radiation reaction across all atom-boundary distances, irrespective of the atomic polarization. This dominance leads to significant spontaneous excitation of the atom. In contrast, in the small-acceleration limit,  the contributions from vacuum fluctuations and radiation reaction nearly cancel each other, resulting in a negligible net rate of change of the mean atomic energy. This balance is observed for atoms polarizable in all directions.  
Moreover, in the intermediate region under large acceleration,  the leading-order contributions from different atomic polarizations are all induced by acceleration and are of the same order of magnitude.  This contrasts sharply with the inertial case, where the dominant contribution arises from polarization normal to the boundary. This shift reflects the profound modification of atomic radiative properties due to large centripetal acceleration.

The combined effect of the boundary and centripetal acceleration plays an important role in modulating the radiative properties of atoms. In both the intermediate and far regions (with the exception of excited atoms under small acceleration in the intermediate regime), the boundary-related terms arise from this combined effect. 
These terms  become the leading contribution in the intermediate region and subleading in the far region, indicating that the boundary may significantly affect atomic transitions even at large separations. The interplay between the boundary and centripetal acceleration also gives rise to a number of interesting behaviors:  an acceleration-independent term  emerges in the radiation reaction contribution  in the intermediate region under large acceleration. 
Moreover, in the far region, the correction to the rate of change of the mean atomic energy arising from the combined effect of the boundary and acceleration decreases monotonically with increasing atom-boundary distance under large acceleration, while under small acceleration, it exhibits an oscillatory decay.

Our results provide new insights into how centripetal acceleration, boundary conditions, and atomic polarization collectively shape the radiative properties of atoms. In particular, we show that a reflecting boundary introduces substantial and sometimes dominant corrections to the atomic transition dynamics, even at large separations.  These results deepen our understanding of quantum field behavior in non-inertial frames and may have implications for experimental studies of boundary-modified radiative effects and the circular Unruh effect.

\section*{Acknowledgments}
We would like to thank the anonymous referee and Wenting Zhou for the constructive suggestions. This work was supported in part by the NSFC under Grant No. 12075084, and the innovative research group of Hunan Province under Grant No. 2024JJ1006.

\section*{Date availability}
No data were created or analyzed in this.

\appendix
\begin{widetext}
\section{CORRELATION FUNCTIONS OF THE FLUCTUATING ELECTROMAGNETIC FIELDS}\label{correlation-function}

The electric field $E_i$ corresponds to the $i0$ component of the electromagnetic tensor $F_{\mu\nu}$, which is defined in terms of the vector potential $A_\mu$ as $F_{\mu\nu}=\partial_\mu A_\nu - \partial_\nu A_\mu$. Explicitly, the electromagnetic tensor $F_{\mu\nu}$ can be represented in matrix form as
\beq
F_{\mu\nu}=
\left(
  \begin{array}{cccc}
     0           & -E_x        & -E_y     & -E_z\\
     E_x         & 0           & B_z      & -B_y\\
     E_y         & -B_z        & 0        & B_x\\
     E_z         & B_y         & -B_x     & 0
  \end{array}
\right).
\eeq

After a Lorentz transformation followed by a rotational transformation, the electric field $E'_i~(i=\rho,\phi,z)$ in the proper frame of the atom expressed in cylindrical coordinates can be written as $E'_i=S_{i}^j(\tau)\Lambda_j^\mu(\tau)\Lambda_0^\nu(\tau) F_{\mu\nu}(x(\tau))$, where 
\begin{equation}
S_{i}^j=\left(
  \begin{array}{ccc}
    \cos(\Omega\gamma\tau) & -\sin(\Omega\gamma\tau) & 0\\
    \sin(\Omega\gamma\tau) &  \cos(\Omega\gamma\tau) & 0\\
       0                   &       0                 & 1
  \end{array}
\right),
\end{equation}
is the rotation matrix and 
\beq\label{Ax}
\Lambda_\iota^\mu=\left(
  \begin{array}{cccc}
    \gamma           & -\gamma\beta n_x  & -\gamma\beta n_y  & 0\\
    -\gamma\beta n_x & 1+(\gamma-1)n_x^2 & (\gamma-1)n_x n_y & 0\\
    -\gamma\beta n_y & (\gamma-1)n_y n_x & 1+(\gamma-1)n_y^2 & 0\\
    0                &         0         &       0           & 1
  \end{array}
\right).
\eeq
is the Lorentz transformation matrix \cite{MTW}, with $n_x=v_x/v=-\sin(\Omega\gamma\tau)$, $n_y=v_y/v=\cos(\Omega\gamma \tau)$ being the unit vector of the velocity,  $\gamma=1/\sqrt{1-\beta^2}$ being the Lorentz factor, and $\beta=v=\Omega R$ being in the natural units.  
Then, the two-point functions in the proper frame of the atom described in cylindrical coordinates can be expressed as \cite{She19}
\begin{eqnarray}
  G_{ij}(u)=\langle 0| E'_i(x(\tau)) E'_k(x(\tau')) |0\rangle
=\langle 0| S_{i}^j(\tau)\Lambda_j^\mu(\tau)\Lambda_0^\nu(\tau) F_{\mu\nu}(x(\tau))\, S_{k}^m(\tau')\Lambda_m^\alpha(\tau')\Lambda_0^\epsilon(\tau') F_{\alpha\epsilon} (x(\tau')) |0\rangle\;,
\end{eqnarray}
which are invariant under temporal translations, so they are functions of $u=\tau-\tau'$. 
Consequently, the two-point functions of the electric field in the proper frame of the atom can be related to the two-point function of the vector potential $\langle0| A_\mu(x(\tau)) A_{\nu}(x(\tau'))|0\rangle$ in the laboratory frame, whose explicit expression is
\begin{align}
&\langle0| A_\mu(x(\tau)) A_{\nu}(x(\tau'))|0\rangle=\nonumber\\
&\frac{1}{4 \pi^2}\left[\frac{\eta_{\mu\nu}}{{
(x-x^\prime)^2+(y-y^\prime)^2+(z-z^\prime)^2-(t-t^\prime-i\varepsilon)^2}}-{\frac{\eta_{\mu\nu}-2n_\mu n_\nu}{
(x-x^\prime)^2+(y-y^\prime)^2+(z+z^\prime)^2-(t-t^\prime-i\varepsilon)^2}}\right]\;,
\end{align}
where $\eta_{\mu\nu}={\rm diag}(-1,1,1,1)$ and $n_{\mu}=(0,0,0,1)$. Finally, the two-point functions in the cylindrical coordinate frame $G_{ij}(u)$, $(i,j=\rho,\phi,z)$ can be obtained as
\begin{align}
G_{zz}(u) =&\frac{\gamma^2\{2R^2+\gamma^2 u^2+2R^4\Omega^2+R^2[(-2-2R^2\Omega^2+\gamma^2 u^2\Omega^2)\cos h-4h \sin h]\}}{\pi^2(-2R^2+\gamma^2 u^2+2R^2\cos h)^3}\nonumber\\
&+\frac{\gamma ^2 R^2 \{\cos h [2-\Omega ^2 \left(4 L^2-2 R^2+\gamma ^2 u^2\right)]+4 h \sin h \}}{\pi ^2 \left(-2 R^2 \cos h+4 L^2+2 R^2-\gamma ^2 u^2\right)^3}+\frac{\gamma ^2 \left(4 L^2-2R^2 \left(R^2 \Omega ^2+1\right)-\gamma ^2 u^2\right)}{\pi ^2 \left(-2 R^2 \cos h+4 L^2+2 R^2-\gamma ^2 u^2\right)^3}\label{Gzz} ,\\
G_{\phi\phi}(u)=&\frac{-2R^2+(2R^2+u^2\gamma^2)\cos h}{\pi^2(-2R^2+u^2\gamma^2+2R^2\cos h)^3}+\frac{\cos h \left(4 L^2+2 R^2+\gamma ^2 u^2\right)-2 R^2}{\pi ^2 \left(-2 R^2 \cos h+4 L^2+2 R^2-\gamma ^2 u^2\right)^3}\label{Gvarphivarphi},\\
G_{\rho\rho}(u)=&\frac{\gamma^2\{[u^2\gamma^2-2(R^2+R^4\Omega^2)]\cos h+R^2[2+(2R^2+u^2\gamma^2)\Omega^2-4h \sin h]\}}{\pi^2(-2R^2+u^2\gamma^2+2R^2\cos h)^3}\nonumber\\
&+\frac{\gamma ^2 R^2 [-4 h \sin h+\Omega ^2 \left(-4 L^2+2 R^2+\gamma ^2 u^2\right)+2] }{\pi ^2 \left(-2 R^2 \cos h+4 L^2+2 R^2-\gamma ^2 u^2\right)^3}+\frac{\gamma ^2 \cos h [4 L^2-2 \left(R^4 \Omega ^2+R^2\right)+\gamma ^2 u^2] }{\pi ^2 \left(-2 R^2 \cos h+4 L^2+2 R^2-\gamma ^2 u^2\right)^3}\label{Grhorho},\\
G_{\rho\phi}(u)=&-G_{\phi\rho}(u)
=\frac{u\gamma^2[2R^2\Omega(-1+\cos h)+u\gamma\sin h]}{\pi^2(-2R^2+u^2\gamma^2+2R^2\cos h)^3}+\frac{\gamma  \left(\sin h \left(4 L^2+\gamma ^2 u^2\right)+2 h R^2 (\cos h-1)\right)}{\pi ^2 \left(-2 R^2 \cos h+4 L^2+2 R^2-\gamma ^2 u^2\right)^3}\label{Grhophi},\\
G_{\phi z}(u)=&-G_{ z \phi}(u)=\frac{4 \gamma  L R (\sin h-h  \cos h)}{\pi ^2 \left(-2 R^2 \cos h+4 L^2+2 R^2-\gamma ^2 u^2\right)^3}\label{Gzvarphi},
\end{align}
and
\begin{align}
G_{z\rho}(u)=&G_{\rho z}(u)=\frac{4 \gamma ^2 L R [(1-\cos h) \left(R^2 \Omega ^2+1\right)-h\sin h]}{\pi ^2 \left(-2 R^2 \cos h+4 L^2+2 R^2-\gamma ^2 u^2\right)^3}\label{Gzrho},~~~~~~~~~~~~~~~~~~~~~~~~~~~~~~~~~~~~~~~~~~~~~~~~~~~~~~~~~
\end{align} 
where $h=u\gamma\Omega$ is defined for brevity.

For centripetally accelerated atoms, the radius $R$ and the rotational angular velocity $\Omega$ can be expressed with the velocity $v$ and centripetal acceleration $a$ as $R=\gamma^2v^2/a$ and $\Omega=v/R=a/\gamma^2v$. Substituting the above two equations and the Lorentz factor into Eqs.~\eqref{Gzz}$\sim$\eqref{Gzrho}, and 
taking the ultrarelativistic limit  $v \rightarrow 1$, the two-point functions become
\begin{align}
G_{zz}=&\frac{24(72+6a^2u^2+a^4u^4)}{\pi^2u^4(12+a^2u^2)^3}+\frac{24 \left(u^2 \left(a^4 u^4+6 a^2 u^2+72\right)-144 L^2 \left(a^2 u^2+2\right)\right)}{\pi ^2 \left(a^2 u^4-48 L^2+12 u^2\right)^3}\label{Gzz1},\\
G_{\phi\phi}=&-\frac{144(12-5a^2u^2)}{\pi^2u^4(12+a^2u^2)^3}+\frac{144 \left(-5 a^2 u^4+48 L^2+12 u^2\right)}{\pi ^2 \left(48 L^2-u^2 \left(a^2 u^2+12\right)\right)^3}\label{Gvarphivarphi1},\\
G_{\rho\rho}=&\frac{24(72-30a^2u^2+a^4u^4)}{\pi^2u^4(12+a^2u^2)^3}-\frac{24 \left(u^2 \left(a^4 u^4-30 a^2 u^2+72\right)-144 L^2 \left(a^2 u^2-2\right)\right)}{\pi ^2 \left(a^2 u^4-48 L^2+12 u^2\right)^3}\label{Grhorho1},\\
G_{\rho\phi}=&-G_{\phi\rho}=-\frac{144a(-12+a^2u^2)}{\pi^2u^3(12+a^2u^2)^3}+\frac{144 a u \left(a^2 u^4-48 L^2-12 u^2\right)}{\pi ^2 \left(a^2 u^4-48 L^2+12 u^2\right)^3}\label{Grhovarphi1},\\
G_{\phi z}=&-G_{z \phi}=\frac{2304 a^2 L u^3}{\pi ^2 \left(48 L^2-u^2 \left(a^2 u^2+12\right)\right)^3}\label{Gzvarphi1},
\end{align}
and
\begin{align}
G_{z\rho}=&G_{\rho z}=\frac{576 a L u^2 \left(a^2 u^2-6\right)}{\pi ^2 \left(48 L^2-u^2 \left(a^2 u^2+12\right)\right)^3}\label{zrho1}.~~~~~~~~~~~~~~~~~~~~~~~~~~~~~~~~~~~~~~~~~~~~~~~~~~~~~~
\end{align}
Note that since $G_{\rho\phi}=-G_{\phi\rho}$ and $G_{\phi z}=-G_{z \phi}$, these cross-terms cancel each other out, resulting in a net contribution of zero to the rate of change of the mean atomic energy.

\section{RATE OF CHANGE OF THE MEAN ATOMIC ENERGY FOR CENTRIPETALLY ACCELERATED ATOMS IN FREE SPACE} \label{free-average-rate}
By integrating Eqs.~\eqref{ijVF} and \eqref{ijRR} and taking the limit where the atom-boundary distance tends to infinity (i.e., $L\to \infty$), one recovers the results for centripetally accelerated atoms in free space. In this limit, the contributions of vacuum fluctuations and radiation reaction to the rate of change of the mean atomic energy are given by
\begin{align}
{\biggl\langle {\frac{dH_A(\tau)}{d\tau}}\biggr\rangle}_{\rm VF}^{\rm free} \approx
&{-\frac{ e^2}{3\pi}
} \left(\sum_{\omega_b>\omega_d}-\sum_{\omega_b<\omega_d}\right)\omega_{bd}^4
\,\bigg \{\frac{\bar{\mathcal{R}}_{\rho \rho}+\bar{\mathcal{R}}_{\phi\phi }+\bar{\mathcal{R}}_{zz}}{2} + \frac{a^2 \left(4 \bar{\mathcal{R}}_{\rho \rho}+4\bar{\mathcal{R}}_{\phi\phi }+\bar{\mathcal{R}}_{zz}\right)}{2 |\omega_{bd}|^2} +e^{-\frac{2 \sqrt{3} |\omega_{bd}|}{a}}\times\nonumber\\
&\left[\frac{\sqrt{3} a \left(4 \bar{\mathcal{R}}_{\rho \rho}+3 \bar{\mathcal{R}}_{\phi\phi }+\bar{\mathcal{R}}_{zz}\right)}{8 |\omega_{bd}|}+\frac{a^2 \left(26 \bar{\mathcal{R}}_{\rho \rho}+23 \bar{\mathcal{R}}_{\phi\phi }+5 \bar{\mathcal{R}}_{zz}\right)}{16 |\omega_{bd}|^2}+\frac{a^3 \left(58 \bar{\mathcal{R}}_{\rho \rho}+55 \bar{\mathcal{R}}_{\phi\phi }+13 \bar{\mathcal{R}}_{zz}\right)}{32 \sqrt{3} |\omega_{bd}|^3}\right]\;\bigg \}\;\label{free-space-vf}
\end{align}
and
\begin{align}
{\biggl\langle {\frac{dH_A(\tau)}{d\tau}}\biggr\rangle}_{\rm RR}^{\rm free} \approx
&{-\frac{ e^2}{3\pi}
}\left(\sum_{\omega_b>\omega_d}+\sum_{\omega_b<\omega_d}\right)\omega_{bd}^4
\bigg [\frac{\bar{\mathcal{R}}_{\rho \rho}+\bar{\mathcal{R}}_{\phi\phi }+\bar{\mathcal{R}}_{zz}}{2}+\frac{a^2 \left(4 \bar{\mathcal{R}}_{\rho \rho}+4\bar{\mathcal{R}}_{\phi\phi }+\bar{\mathcal{R}}_{zz}\right)}{2 |\omega _{bd}|^2} \bigg]\;,\label{free-space-rr}
\end{align}
respectively. Summing these two contributions gives the total rate change of the mean atomic energy, which is expressed as 
\begin{align}
{\biggl\langle {\frac{dH_A(\tau)}{d\tau}}\biggr\rangle}_{\rm tot} \approx
&{-\frac{ e^2}{3\pi}
} \sum_{\omega_b>\omega_d}\omega_{bd}^4
\,\bigg \{\bar{\mathcal{R}}_{\rho \rho}+\bar{\mathcal{R}}_{\phi\phi }+\bar{\mathcal{R}}_{zz}+ \frac{a^2 \left(4 \bar{\mathcal{R}}_{\rho \rho}+4\bar{\mathcal{R}}_{\phi\phi }+\bar{\mathcal{R}}_{zz}\right)}{ |\omega_{bd}|^2} +e^{-\frac{2 \sqrt{3} |\omega_{bd}|}{a}}\times\nonumber\\
&\left[\frac{\sqrt{3} a \left(4 \bar{\mathcal{R}}_{\rho \rho}+3 \bar{\mathcal{R}}_{\phi\phi }+\bar{\mathcal{R}}_{zz}\right)}{8 |\omega_{bd}|}+\frac{a^2 \left(26 \bar{\mathcal{R}}_{\rho \rho}+23 \bar{\mathcal{R}}_{\phi\phi }+5 \bar{\mathcal{R}}_{zz}\right)}{16 |\omega_{bd}|^2}+\frac{a^3 \left(58 \bar{\mathcal{R}}_{\rho \rho}+55 \bar{\mathcal{R}}_{\phi\phi }+13 \bar{\mathcal{R}}_{zz}\right)}{32 \sqrt{3} |\omega_{bd}|^3}\right]\;\bigg \}\nonumber\\
&+\frac{ e^2}{3\pi}\sum_{\omega_b<\omega_d} \omega_{bd}^4
\, \bigg \{ e^{-\frac{2 \sqrt{3} |\omega_{bd}|}{a}}\left[\frac{\sqrt{3} a \left(4 \bar{\mathcal{R}}_{\rho \rho}+3 \bar{\mathcal{R}}_{\phi\phi }+\bar{\mathcal{R}}_{zz}\right)}{8 |\omega_{bd}|}+\frac{a^2 \left(26 \bar{\mathcal{R}}_{\rho \rho}+23 \bar{\mathcal{R}}_{\phi\phi }+5 \bar{\mathcal{R}}_{zz}\right)}{16 |\omega_{bd}|^2}\right.\nonumber\\
&\left.+\frac{a^3 \left(58 \bar{\mathcal{R}}_{\rho \rho}+55 \bar{\mathcal{R}}_{\phi\phi }+13 \bar{\mathcal{R}}_{zz}\right)}{32 \sqrt{3} |\omega_{bd}|^3}\right]\;\bigg \}\;.\label{free-space-tot}
\end{align}

In the limit of large acceleration (i.e., $a/ |\omega_{bd}|\gg 1$), the contributions of vacuum fluctuations and radiation reaction to the rate of change of the mean atomic energy take the following forms:
\begin{align}
{\biggl\langle {\frac{dH_A(\tau)}{d\tau}}\biggr\rangle}_{\rm VF}^{\rm free} \approx
&{-\frac{ e^2}{3\pi}
} \left(\sum_{\omega_b>\omega_d}-\sum_{\omega_b<\omega_d}\right)\omega_{bd}^4
\,\bigg [\frac{a^3 \left(58 \bar{\mathcal{R}}_{\rho \rho}+55 \bar{\mathcal{R}}_{\phi\phi }+13 \bar{\mathcal{R}}_{zz}\right)}{32 \sqrt{3} |\omega _{bd}|^3}+\frac{\sqrt{3} a \left(14 \bar{\mathcal{R}}_{\rho \rho}+15 \bar{\mathcal{R}}_{\phi\phi }+5 \bar{\mathcal{R}}_{zz}\right)}{16 |\omega _{bd}|}\;\bigg ]\;\label{freevf}
\end{align}
and
\begin{align}
{\biggl\langle {\frac{dH_A(\tau)}{d\tau}}\biggr\rangle}_{\rm RR}^{\rm free} \approx
&{-\frac{ e^2}{3\pi}
} \left(\sum_{\omega_b>\omega_d}+\sum_{\omega_b<\omega_d}\right)\omega_{bd}^4
\,\bigg [\frac{a^2  \left(4 \bar{\mathcal{R}}_{\rho \rho}+4\bar{\mathcal{R}}_{\phi\phi }+\bar{\mathcal{R}}_{zz}\right)}{2 | \omega _{bd}|^2}+\frac{\bar{\mathcal{R}}_{\rho \rho}+\bar{\mathcal{R}}_{\phi\phi }+\bar{\mathcal{R}}_{zz}}{2}\; \bigg ]\;.\label{freerr}
\end{align}

In the limit of small acceleration (i.e., $a/\omega_{bd} \ll 1$), the contributions of vacuum fluctuations and radiation reaction to the rate of change of the mean atomic energy for centripetally accelerated atoms in free space can still be expressed as Eqs. \eqref{free-space-vf} and \eqref{free-space-rr}, respectively. For atoms in the excited state, the leading term due to acceleration is proportional to $a^2$. In contrast, for atoms in the ground state, the rate of change is exponentially suppressed.

\end{widetext}

\end{document}